\documentclass[usenatbib,twocolumn]{aastex63}

\usepackage{amsmath}
\usepackage{graphicx}
\received{\today}

\submitjournal{ApJ}
\shorttitle{Modelling the dust in SN 2010jl}
\shortauthors{Bevan et al.}

\begin{document}

\title{Disentangling Dust Components in SN 2010jl: The First 1400 Days}

\correspondingauthor{A. M. Bevan}
\email{antonia.bevan.12@ucl.ac.uk}

\author{A. M. Bevan}
\affiliation{Department of Physics and Astronomy, University College London, Gower Street, London WC1E 6BT, UK}

\author{K. Krafton}
\affiliation{Department of Physics \& Astronomy, Louisiana State University, Baton Rouge, LA 70803, USA}

\author{R. Wesson}
\affiliation{Department of Physics and Astronomy, University College London, Gower Street, London WC1E 6BT, UK}

\author{J. E. Andrews}
\affiliation{Steward Observatory, University of Arizona, 933 North Cherry Avenue, Tucson, AZ 85721, USA}

\author{E. Montiel}
\affiliation{SOFIA-USRA, NASA Ames Research Center, MS 232-12, Moffett Field, CA 94035, USA}

\author{M. Niculescu-Duvaz}
\affiliation{Department of Physics and Astronomy, University College London, Gower Street, London WC1E 6BT, UK}

\author{M. J. Barlow}
\affiliation{Department of Physics and Astronomy, University College London, Gower Street, London WC1E 6BT, UK}

\author{I. De Looze}
\affiliation{Department of Physics and Astronomy, Ghent University, Krijgslaan 281, S9, 9000 Gent, Belgium}

\author{G. C. Clayton}
\affiliation{Department of Physics \& Astronomy, Louisiana State University, Baton Rouge, LA 70803, USA}

\begin{abstract}
	The luminous Type IIn SN 2010jl shows strong signs of interaction between the SN ejecta and dense circumstellar material. Dust may be present in the unshocked ejecta, the cool, dense shell between the shocks in the interaction region, or in the circumstellar medium. We present and model new optical and infrared photometry and spectroscopy of SN 2010jl from 82 to 1367 days since explosion. We evaluate the photometric and spectroscopic evolution using the radiative transfer codes {\sc{mocassin}} and {\sc{damocles}}, respectively. We propose an interaction scenario and investigate the resulting dust formation scenarios and dust masses. We find that SN 2010jl has been continuously forming dust based on the evolution of its infrared emission and optical spectra. There is evidence for pre-existing dust in the circumstellar medium as well as new dust formation in the cool, dense shell and/or ejecta. We estimate that 0.005--0.01~M$_{\sun}$ of predominantly carbon dust grains has formed in SN 2010jl by  $\sim$1400\,d post-outburst. 
\end{abstract}


\section{introduction}
\label{sec:intro}

Dust evolution in Type IIn supernovae (SNe) is complex. Non-interacting SNe begin forming dust in their cooling ejecta generally several hundred days post-outburst \citep[e.g.][]{2006Sci...313..196S,2017MNRAS.465.4044B,2017MNRAS.469.3347M}. However, in a Type IIn SN, the interaction between the forward shock and the surrounding, dense circumstellar medium (CSM) at much earlier times propagates a reverse shock back into the ejecta complicating this picture. Between the two shocks is a rapidly cooling region, the cool, dense shell (CDS), in which new grains of dust may be able to condense \citep{2004MNRAS.352..457P}. Dust grains may still be able to condense in the unshocked ejecta and pre-existing dust that formed during pre-supernova mass loss phases may also be present in the CSM. Disentangling the contributions of each of these dust components to the observable signatures of interacting SNe is important for understanding the evolution of dust in interacting SNe as well as for interpreting their observations. 

The bright Type IIn SN\,2010jl is an interesting laboratory for the study of dust formation, evolution, and destruction. SN 2010jl was discovered on 2010 November 3.52 in the irregular galaxy UGC 5189A at a distance of 48.9 $\pm$ 3.4 Mpc \citep{2010CBET.2532....1N} and following its discovery has been well-observed. With a peak absolute magnitude of $\sim$-20, this was one of the brightest SNe in recent years \citep{2011ApJ...730...34S}. Early spectra of its broad emission features near maximum light showed narrow line emission, leading to SN 2010jl's classification as a Type IIn SN  \citep{2010CBET.2536....1B,2010CBET.2539....1Y}. SN\,2010jl's interaction with dense circumstellar material (CSM) allowed it to be detected in X-rays \citep{2010ATel.3012....1I} and it is among the most luminous X-ray SNe yet observed \citep{2012ApJ...750L...2C}.  

The often long-lasting circumstellar interaction of Type IIn SNe results in a significantly more complex geometry than non-interacting supernovae.  Generally, interacting supernovae are assumed to consist of concentric shells of unshocked ejecta, reverse-shocked ejecta adjoining forward-shocked CSM with a cool, dense shell of material at their boundary, and unshocked CSM outside of the forward shock. Diagrams illustrating this scenario are abundant in the literature \citep[see e.g.][]{1994ApJ...420..268C,2017hsn..book..403S,2018ApJ...859...66S}. The characteristic narrow line emission seen in Type IIn SNe arises in the flash-ionized CSM. In the case of SN 2010jl, there are numerous narrow optical emission features. In most cases, these displayed a P-Cygni proile for the first few hundred days indicative of high densities in the line-emitting region of the CSM \citep{2014ApJ...797..118F}.  Pre-existing dust may be present in the unshocked CSM having formed in pre-supernova outbursts or winds.  The presence of an IR-excess through the first $\sim400$ days of the evolution of SN 2010jl has been attributed to a light echo caused by such pre-existing, circumstellar dust \citep{2011MNRAS.413L.101K,2014ApJ...797..118F}. 

In addition to pre-existing, circumstellar dust, SN2010jl exhibits signatures of newly-formed dust associated with the interacting system. New dust formation inside core-collapse supernovae (CCSNe) can give rise to three observable signatures: (1) an IR excess due to thermal emission from hot or warm dust, (2) a concurrent increased rate of fading of the optical flux, and (3) a progressive and systematic blueshift of emission line profiles as the receding parts of the expanding SN are increasingly blocked by new dust. The IR excess can also be caused by pre-existing circumstellar dust heated by the UV flash, but pre-existing dust likely only makes a significant contribution to the IR excess at early times since it fades rapidly for most geometries. SN 2010jl exhibited an IR excess at early and later stages along with blue-shifted line profiles, leading a number of authors to infer the formation of new dust in its CDS and/or in its ejecta \citep[e.g.][]{2011AJ....142...45A,2012AJ....143...17S,2013ApJ...776....5M,2014Natur.511..326G,2014ApJ...797..118F,2018MNRAS.481.3643C,2018ApJ...859...66S}.  

In this paper, we have determined the evolution of dust in SN 2010jl over the first 1400 days. We have further sought to disentangle pre-existing dust in the unshocked CSM from newly-formed dust behind the SN shock front. We have obtained new visible and IR photometry and spectroscopy using Gemini/GMOS and {\it Spitzer}/IRAC during the first 1400 days of SN 2010jl's evolution which we have supplemented with data from online archives. We have built a careful model of SN2010jl based on these data and have applied two 3D Monte Carlo radiative transfer codes ({\sc mocassin} and {\sc damocles}) to simultaneously model the spectral energy distributions (SEDs) and asymmetric dust-affected emission line profiles at a range of epochs in order to distinguish between different dust populations in SN2010jl and to determine their respective dust masses. 

This paper is organized as follows. In Section \ref{sec:obs}, we describe the optical and infrared photometry and optical spectroscopy obtained with the Gemini North and South telescopes, the {\it {Spitzer Space Telescope}}, and the European Southern Observatory's Very Large Telescope and New Technology Telescope (ESO/VLT/NTT). In Section \ref{sec:dust}, we discuss observational signatures of dust in SN 2010jl before presenting our {\sc damocles} modeling of red-blue asymmetries in the optical emission lines and {\sc mocassin} modeling of optical-IR spectral energy distributions (SEDs) in Section \ref{sec:models}. In Section \ref{sec:discussion}, we present a discussion of our results before summarising in Section \ref{sec:concs}.

\section{Observations}
 \label{sec:obs}
The first detection of SN 2010jl was from pre-discovery images obtained on 2010 October 9.6 \citep{2011ApJ...730...34S}. For this paper, we adopt an explosion date of 2010 October 10. 

\subsection{Photometry}

New images and spectra of SN 2010jl were obtained over 5 epochs with Gemini/GMOS-S (GS-2012A-Q-79, GS-2013A-Q-93, GS-2014A-Q-70, GN-2016A-Q-85-31). The g$^{\prime}$r$^{\prime}$i$^{\prime}$ images were reduced and stacked using the IRAF 
\textit{gemini} package \citep{Tody1986,Tody1993}.  The instrumental g$^{\prime}$r$^{\prime}$i$^{\prime}$  magnitudes were transformed to standard Johnson-Cousins VRI using the tertiary standards presented in \citet{2011AJ....142...45A}, and transformations presented in \citet{2007ApJ...669..525W}. Uncertainties were calculated by adding in quadrature using the transformation uncertainty quoted in \citet{2007ApJ...669..525W}, photon statistics, and the zero point deviation of the standard stars for each epoch. The derived VRI magnitudes are provided in Table \ref{tb_photo_BVRI}. Differences between our VRI magnitudes and those presented by \citet{2014ApJ...797..118F}  may be due to their corrections to account for line emission. We did not apply any such corrections to our data as, except for the R-band, we do not find that line emission significantly contributes to the broad-band fluxes.

{\it Spitzer}/IRAC (3.6 and 4.5 $\mu$m) images of SN 2010jl were obtained at six epochs during 2011-2014. A further five epochs during this period were taken from the Spitzer Heritage Archive\footnote{https://sha.ipac.caltech.edu/applications/Spitzer/SHA/} \citep{2013AJ....146....2F,2019ApJS..241...38S}, along with the pipeline basic calibrated data. Pre-explosion IRAC images of UGC 5189A, available in the {\it Spitzer} archive from 2007 December 27 (Program 40301, PI Fazio), were used to subtract the host galaxy from the SN 2010jl images to allow for more accurate photometry. Aperture photometry was performed using standard IRAF routines (digiphot/apphot/phot). 
Eight IRAC observations were obtained with frametimes of 100\,s and three of 30\,s. Some of the 100 s integrations suffer from saturation effects in some pixels.  However, at three epochs (253\,d, 993\,d and 1367\,d),  unsaturated observations with 30\,s exposures were also obtained within a few days of the 100\,s observations. The 30\,s exposures were used to correct the fluxes of the 100 s observations for the saturation effects.
The 30\,s and 100\,s  exposures taken on days 253 and 260, respectively, showed only a 2\% difference in the count rates, which we considered to be negligible. At days 993 and 997, however, the 30\,s exposure had a significantly higher count rate than the 100 s exposure in both bands, by 38\% in the 3.6 $\mu$m band and 31\% in the 4.5 $\mu$m band. On days 1367-8, the 30\,s exposure was brighter than two 100\,s exposures in both bands, by 13\% in the 3.6 $\mu$m band and 12\% in the 4.5 $\mu$m band. 

100\,s exposures taken on days 464, 620 and 843 did not have 30\,s exposures taken at similar times. These observations were only partially saturated and we therefore conservatively corrected the flux at these epochs using the 38\% (31\%) difference between the different frametimes in the 3.5$\mu$m (4.6$\mu$m) band at 993\,d as a correction factor. We show the saturated fluxes of 8.55\,mJy (3.6$\mu$m) and 8.30\,mJy (4.5$\mu$m) as a lower limit on our plots in Section \ref{sec:mocassin}.

\begin{figure*}
	\centering
	\includegraphics[width=1\textwidth,clip = true, trim = 43 128 79 150]{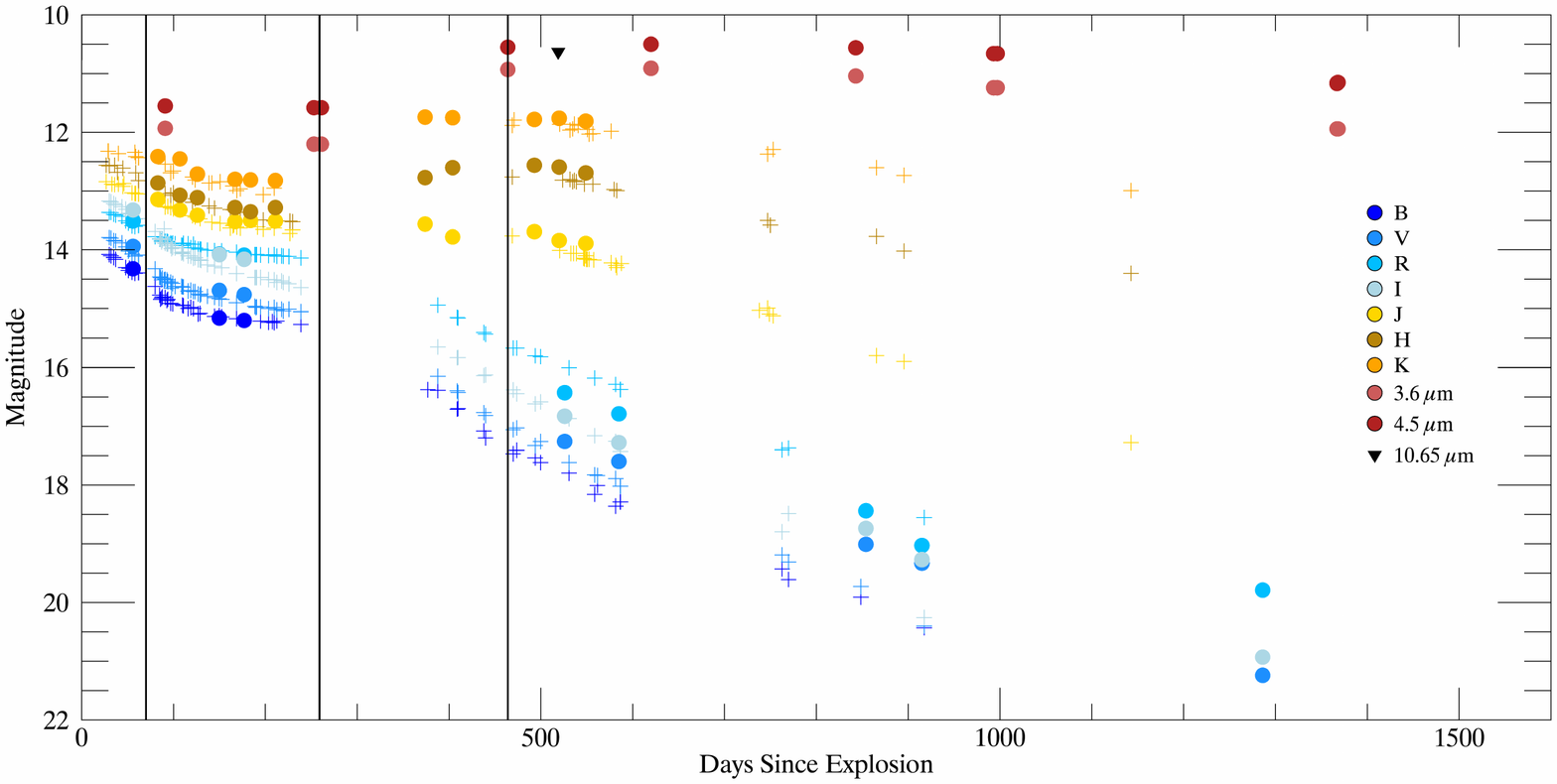} 
		\caption{Optical and IR light-curves of SN 2010jl. Circles represent data from \citet{2011AJ....142...45A} and new data presented in Tables 1-3 in this paper. Crosses represent data from \citet{2014ApJ...797..118F}. The vertical lines represent different landmarks of SN 2010jl's evolution. The first line, at about Day 70, marks a transition in SN 2010jl's spectra. Before this time, the spectra do not have IWCs in any lines. After the first vertical line, IWCs arise in the Balmer and He I $\lambda 5876$ \AA\ lines and begin to shift to the blue. This is also when the initial IR excess emerges. The second vertical line marks the beginning of a gap in the observations. From the second line to the third line there is a sharp rise in the IR fluxes, and the IWCs' blueshifts continue to strengthen.  After the third line, we continue to see evidence of dust formation via an IR excess and blueshifted optical line profiles. }
		\label{Fig:lightcurve}
\end{figure*}

The optical and IR light curves are shown in Figure~\ref{Fig:lightcurve}.
Table \ref{tb_spitzer} lists the {\it Spitzer} fluxes and their 1-$\sigma$ uncertainties for the 7 epochs used for the analysis in this paper. The 100 s observations used to establish the saturation correction are not included in Table  \ref{tb_spitzer}. We estimate that the uncertainty in the saturation correction for individual epochs is $\sim$10\%. These uncertainties were combined with the photometric uncertainties. They are listed in Table \ref{tb_spitzer}. 

\begin{table*}
	\caption{BVRI Photometry of SN 2010jl from Gemini/GMOS-S}
	\def\arraystretch{1.15}
	\centering
	\begin{tabular}{ccccccc}
		\hline
		Date&JD&Age&V&R&I\\
		&&days&mag&mag&mag\\
		\hline
		2012 Mar 19 &2456006&526 &17.26 $\pm$ 0.05&16.43 $\pm$ 0.05 &16.83 $\pm$ 0.02\\
		2012 May 17 &2456065&585 &17.60 $\pm$ 0.06 &16.79 $\pm$ 0.02 &17.28 $\pm$ 0.03\\
		2013 Feb 10 &2456334&854 &19.01 $\pm$ 0.08 &18.44 $\pm$ 0.05&18.74 $\pm$ 0.07\\
		2013 Apr 12 &2456395&915 &19.33 $\pm$ 0.08 &19.03 $\pm$ 0.05&19.27 $\pm$ 0.07\\
		2014 Apr 18 &2456766&1286&21.24 $\pm$ 0.06&19.79 $\pm$ 0.03&20.93 $\pm$ 0.05\\
		\hline
	\end{tabular}
	\label{tb_photo_BVRI}
\end{table*}

JHKs imaging obtained with NTT/SOFI (184.D-1151, PI Benetti) at La Silla was retrieved from the ESO Science Archive Facility.  Reductions were done using the standard procedure in \textsc{iraf}, including crosstalk and flatfield correction, background subtraction, and the shifting and adding of individual image frames.  Aperture photometry was then done using a set of standard stars in the field obtained from the 2MASS catalog. The JHKs photometry is presented in Table \ref{tb_photo_JHK}. 

We observed SN 2010jl with VLT/VISIR on 2012 March 12 (day 519), using the B10.7 filter (10.7 $\mu$m) (ESO program 288.D-5044(A)). The standard VISIR pipeline recipes were used with ESOREX, and flux calibration was done using a standard star observed immediately after the target at similar airmass. The supernova was not detected in this observation; from the root-mean-square variations in the reduced image, we estimated a 3$\sigma$ upper limit of 2~mJy for the 10.7$\mu$m flux density.

\begin{table*}
	\caption{JHKs Photometry of SN 2010jl from NTT/SOFI}
	\def\arraystretch{1.15}
	\centering
	\begin{tabular}{ccccccc}
		\hline
		Date&JD&Age&J&H&K$_{s}$\\
		&&days&mag&mag&mag\\
		\hline
		2011 Jan 1& 2455562&83& 13.10 $\pm$ 0.05&12.75 $\pm$ 0.12&12.35 $\pm$ 0.08\\
		2011 Jan 25&2455587&107&13.25 $\pm$ 0.01&13.01 $\pm$ 0.02&12.46 $\pm$ 0.11\\
		2011 Feb 13&2455606&126&13.30 $\pm$ 0.04&13.04 $\pm$ 0.05&12.65 $\pm$ 0.12\\
		2011 Mar 26&2455647&167&13.46 $\pm$ 0.03&13.21 $\pm$ 0.05&12.75 $\pm$ 0.07\\
		2011 Apr 12&2455664&184&13.43 $\pm$ 0.06&13.25 $\pm$ 0.10&12.85 $\pm$ 0.10\\
		2011 May 9& 2455691&211&13.40 $\pm$ 0.18&13.16 $\pm$ 0.09&12.88 $\pm$ 0.12\\
		2011 Jun 25&2455738&258&13.48 $\pm$ 0.07&13.25 $\pm$ 0.08&12.73 $\pm$ 0.11\\
		2011 Oct 19&2455853&374&13.59 $\pm$ 0.10&12.77 $\pm$ 0.10&11.67 $\pm$ 0.10\\
		2011 Nov 18&2455883&404&13.46 $\pm$ 0.04&12.60 $\pm$ 0.14&11.72 $\pm$ 0.13\\
		2012 Feb 15&2455972&493&13.74 $\pm$ 0.02&12.58 $\pm$ 0.07&11.78 $\pm$ 0.12\\
		2012 Mar 13&2455999&520&13.88 $\pm$ 0.06&12.64 $\pm$ 0.09&11.74 $\pm$ 0.10\\
		2012 Apr 11&2456028&549&13.91 $\pm$ 0.04&12.74 $\pm$ 0.04&11.85 $\pm$ 0.08\\
		\hline
	\end{tabular}
	\centering
	\label{tb_photo_JHK}
\end{table*}

%

\begin{table*}
	\caption{{\it Spitzer}/IRAC Photometry of SN 2010jl. }
	\def\arraystretch{1.15}
	\centering
	\begin{tabular}{llcccc}
		\hline
		Date&JD&Age&3.6 $\mu$m&4.5 $\mu$m\\
		&&days&mJy&mJy\\
		\hline
		2011 Jan 9 &2455571&91&4.14 $\pm$ 0.11&4.32 $\pm$ 0.12\\
		2011 Jun 20 &2455732&253&3.68 $\pm$ 0.10&4.18 $\pm$ 0.12\\
		2012 Jan 17$^a$ &2455944&464&11.75 $\pm$ 0.31&10.84 $\pm$ 0.27\\
		2012 Jun 21$^a$ &2456100&620&12.04 $\pm$ 0.32&11.35 $\pm$0.27 \\
		2013 Jan 30$^a$ &2456323&843&10.62 $\pm$ 0.25&10.69 $\pm$0.25 \\
		2013 Jun 29 &2456472&993&8.81 $\pm$ 0.09&9.81 $\pm$0.10\\
		2014 Jul  8 &2456846&1367&4.67 $\pm$ 0.12& 6.09 $\pm$ 0.16\\
		\hline
	\end{tabular}
	
	$^a$Integrations of 100\,s corrected for saturation effects
	\label{tb_spitzer}
\end{table*}

The extinction from the Milky Way along the line of sight to SN 2010jl is very small, E(B-V) = 0.027 \citep{1998ApJ...500..525S}. The flux measurements are corrected for this Galactic foreground extinction, assuming R$_{V}$ = 3.1 \citep{1989ApJ...345..245C}. There is evidence for a small amount of additional reddening (E(B-V) $\sim$ 0.03) from dust associated with the host galaxy, from the equivalent width measured for a Na~{\sc i} D$_2$ component at the velocity of UGC~5189A \citep{2011A&A...527L...6P}. 

Using our new observations as well as data from the literature \citep{2011AJ....142...45A,2014ApJ...797..118F}, we present optical (BVRI) and IR (JHKs, 3.6, 4.5 \micron) light curves extending out to almost 1400 d in Figure~\ref{Fig:lightcurve}. 

\subsection{Spectroscopy}
For each Gemini/GMOS epoch (526\,d, 585\,d, 854\,d, 915\,d and 1286\,d), spectra were obtained from three 900s exposures taken in longslit mode using the B600 grating and a slit width of 0$\farcs$75. Central wavelengths of 5950, 5970, and 5990 \AA\ were chosen ensure complete spectral coverage accounting for chip gaps. A 2x2 binning in the low gain setting was used. As with the imaging, the spectra were reduced using the IRAF \textit{gemini} package. The sky subtraction regions were determined by visual inspection to prevent contamination from material not associated with the SN, and the spectra were extracted using 15 rows centered on the SN. We also made use of day 34-247 ESO archival Very Large Telescope X-Shooter spectra that were originally presented by \citet{2014Natur.511..326G}. 
The spectra from each individual night have been corrected for the radial velocity of  UGC 5189A (3167 km s$^{-1}$). The spectra and the evolution of the line profiles are presented in Figures~\ref{Fig:optspectra}-\ref{Fig:balmer}.

\section{Observational signatures of dust in SN2010jl}
 \label{sec:dust}
\subsection{The IR excess}
The IR excess, apparent by day 91, persists through all our epochs of observations to day 1367 (Figure~\ref{Fig:lightcurve}). Previous models by \citet{2011AJ....142...45A} and \citet{2014ApJ...797..118F} have suggested that the IR excess emission at early times is likely dominated by an IR echo caused by the flash heating of pre-existing dust in the CSM. At later times, however, it is unlikely that the echo could sustain the observed IR excess (see Section \ref{sec:mocassin}). This points towards a transition in the dominating dust emission component from pre-existing echoing dust in the CSM to warm or hot newly-formed dust in the CDS or ejecta. The near-IR jumps in brightness between days 260-464  at both 3.6 and 4.5 $\mu$m before slowly declining over  a period of nearly 1000\,d. The H and K bands also brighten substantially during this time $\sim$1\,yr post-outburst before steadily declining. The sudden increase in IR flux between days 260-464 is a strong indicator of new dust formation. Concurrently, the rate of decline of the optical bands increases significantly. However, it is not possible to  attribute this accelerated decline definitively to dust formation due to the lack of data between $\sim$200 and 350 days. Conversely, the lack of a wavelength dependence in the rate of decline in the visible bands does not necessarily preclude dust formation being the cause since larger dust grain sizes, as have been inferred to have formed in other CCSNe \citep{2015MNRAS.446.2089W,2017MNRAS.465.4044B}, could reduce the wavelength dependence of extinction in the optical. 

\begin{figure*}
	\centering
	\includegraphics[width=1\textwidth,clip = true, trim = 100 145 50 160]{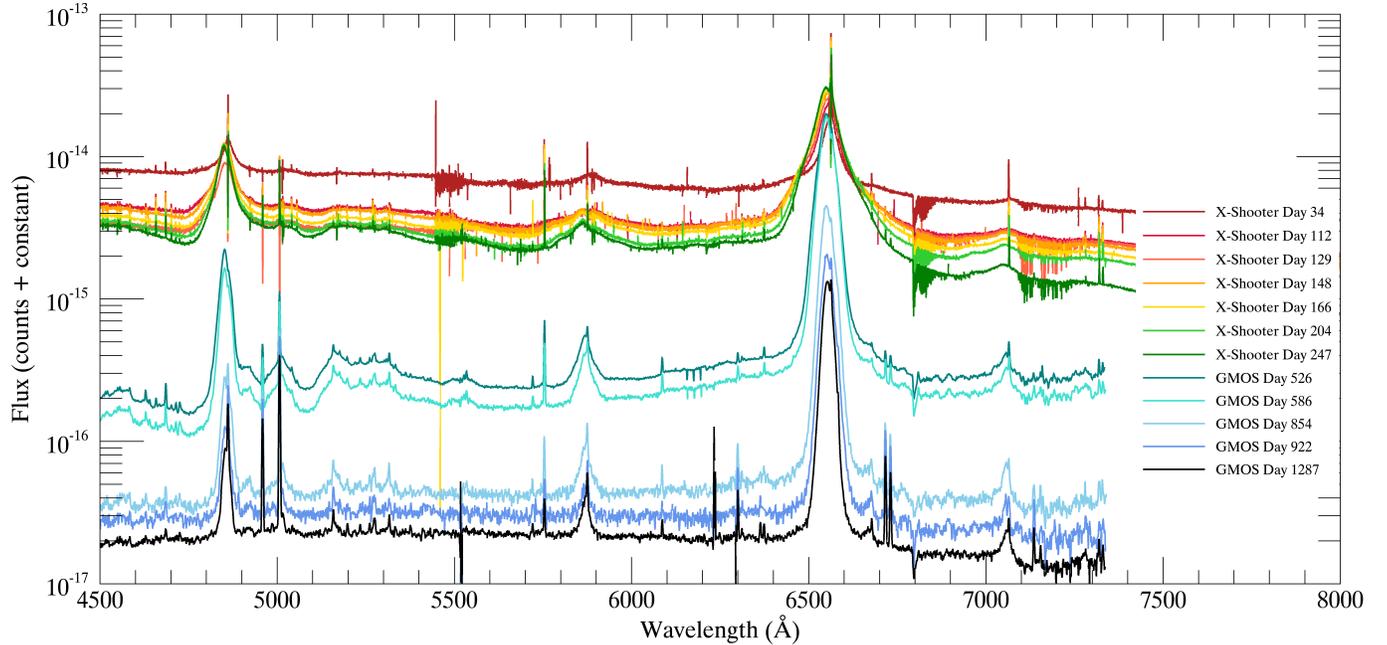}
	\caption{ Optical spectra of SN 2010jl from 34 - 1367\,d post-outburst. Our late-time Gemini/GMOS spectra are shown in addition to VLT/X-Shooter observations from \citet{2014Natur.511..326G}, which were obtained from the ESO archive. All epochs are relative to our assumed explosion date of 2010 October 10. Note that \citet{2014Natur.511..326G} adopt epochs relative to SN~2010jl's peak luminosity on 2010 October 18  and thus differ from those given in this paper.
	}
	\label{Fig:optspectra}
\end{figure*}

\begin{figure*}
	\centering
	\includegraphics[width=0.95\textwidth,clip = true,trim =15 100 15 100]{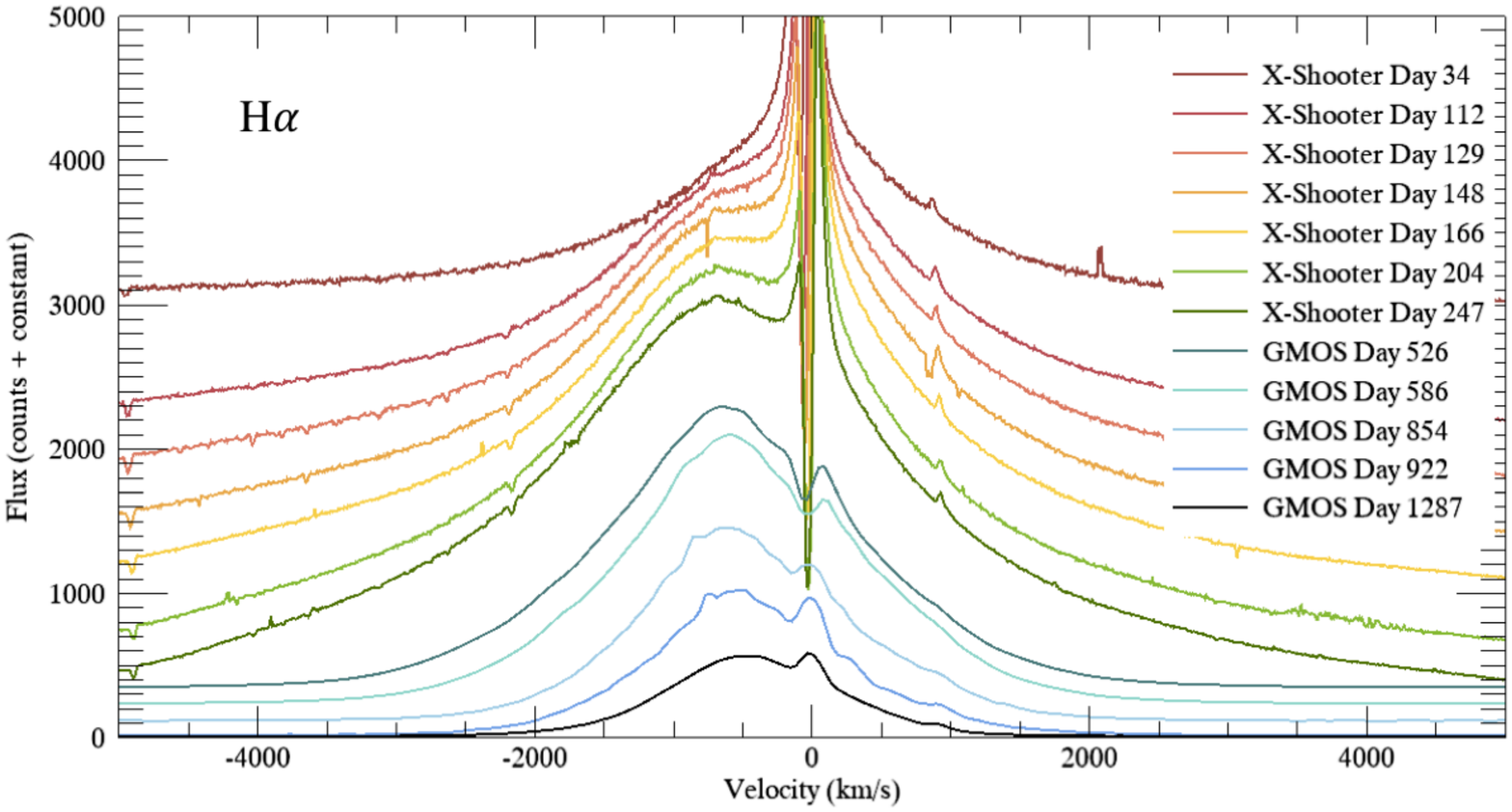} 
	\vspace{0.5cm} \\
	\includegraphics[width=0.45\textwidth,clip = true,trim = 110 20 115 20]{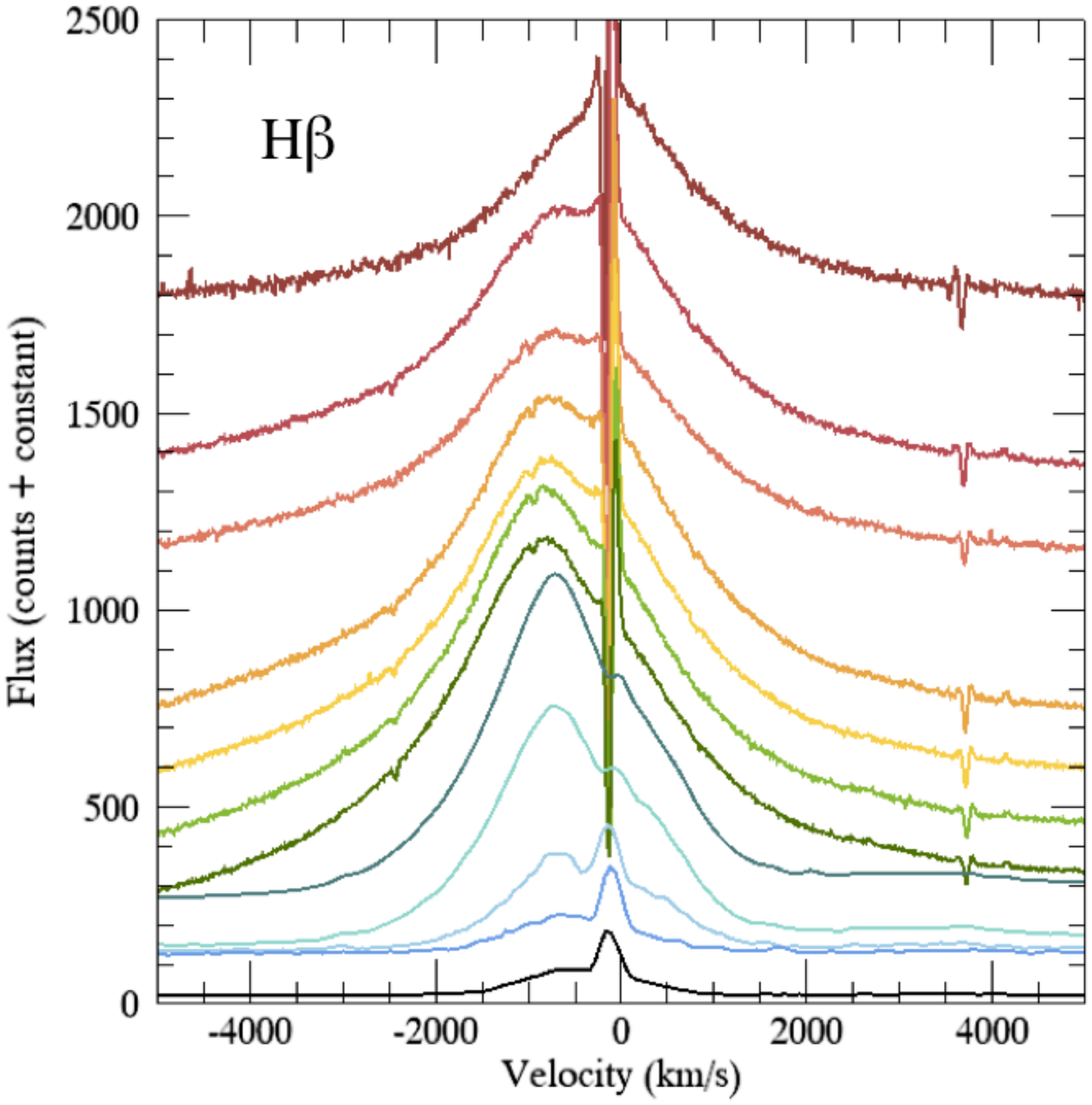} \hspace{0.5cm}
	\includegraphics[width=0.44\textwidth,clip = true,trim = 15 100 25 100]{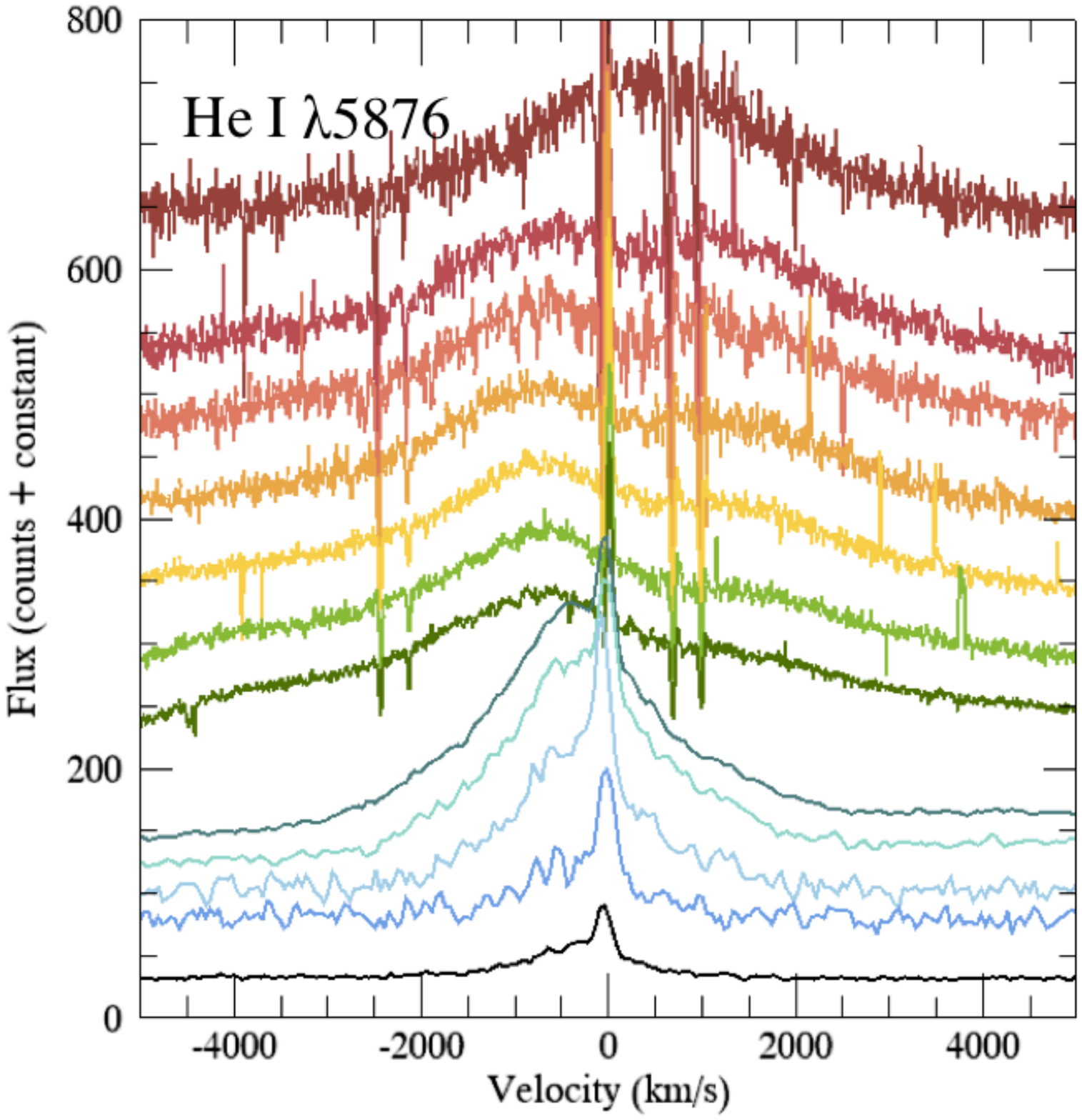} 
	\caption{The evolution of the H$\alpha$ (top), H$\beta$ (left), and He I $\lambda 5876$ \AA (right) line profiles. We attribute the apparent red peak or shoulder of the He I $\lambda 5876$ \AA\ line in the first 3 epochs to another blended line.}
	\label{Fig:balmer}
\end{figure*}

\subsection{Asymmetric emission line profiles}

In order to interpret and model the effects of dust on the emission line profiles of SN 2010jl, we must first consider the different emitting components that contribute to line profiles of interacting SNe. Type IIn SNe are identified by narrow emission lines of $\sim$100 km s$^{-1}$ in width which arise in the flash-ionized, unshocked CSM.  In addition to these narrow width components (NWCs), Type IIn SNe frequently also exhibit intermediate width components (1-4 $\times$ 10$^{3}$ km s$^{-1}$; IWCs) and broad width components (up to $\sim$1 $\times$ 10$^{4}$ km s$^{-1}$; BWCs) \citep[e.g.][]{2009ApJ...695.1334S}. The IWC is likely emitted from a shocked region between the forward and reverse shocks that may consist of accelerated clumps of CSM or of mixed reverse-shocked ejecta and forward-shocked CSM. The BWC could originate from the rapidly expanding ejecta or, at earlier times, may alternatively be the product of electron scattering generating significantly extended wings to the profile. 

The X-Shooter spectrum taken on 2010 November 5, 26 days after outburst, shows that SN 2010jl is a Type IIn SN \citep{2010CBET.2536....1B,2010CBET.2539....1Y}. In this first spectrum, the Balmer-line profiles are apparently Lorentzian in shape with a narrow peak and broad wings ($\sim$15,000 km s$^{-1}$). The narrow width components (NWCs) in SN 2010jl exhibit P Cygni profiles \citep{2011ApJ...732...63S} and are likely sustained at later epochs by X-rays from the shock interaction. 

In the optical spectra, the Balmer series and He I lines exhibit IWCs ($\sim$2000 km s$^{-1}$) from $\sim$80\,d as well as BWCs  (Figures~\ref{Fig:optspectra}-\ref{Fig:balmer}). The IWCs increasingly dominate the total line fluxes over time. During the first 200 days, the line profiles (e.g. H$\alpha$, $H\beta$, He {\sc i} 5876\AA) exhibit a progressive flux bias towards the blue and blueshifted peak fluxes. The blueshifting strengthens through the first 245 days after explosion before receding, much more slowly, back towards zero velocity. The blueshift persists to late epochs (1286\,d).

Such red-blue asymmetries could be caused by a number of effects including high optical depths at early times \citep{2001MNRAS.326.1448C,2012AJ....143...17S,2014ApJ...797..118F}, electron scattering, intrinsic asymmetries in the emissivity distributions or obscuration of the receding side of the supernova by internal dust grains \citep{1989LNP...350..164L}. 

Several other interacting supernovae, such as SN~2005ip, SN~1995N and SN~2006jd, exhibit similar  blueshifted asymmetries in their line profiles \citep[e.g.][Wesson et al. in prep.]{2012ApJ...756..173S, 2019MNRAS.485.5192B}. Whilst an asymmetric geometry could explain the observed line shapes for an individual object, asymmetric geometries are unable to account for both the frequent occurrence of blueshifted emission line profiles in interacting supernovae and the absence of any redshifted equivalents. In addition, the steadily evolving nature of the line asymmetries in SN 2010jl would require a very specific geometry to reproduce the necessary time-dependent emissivity distribution. We therefore disfavour a scenario in which intrinsic geometrical asymmetries are the cause of the asymmetric emission lines in SN~2010jl.

The persistence of asymmetrical lines to very late times ($>$1000\,d) when the photosphere has entirely receded rules out line optical depth effects at late times \citep{2018MNRAS.481.3643C}, and, whilst scattering by electrons in the CSM could cause long-lasting line shifting, it would affect all lines similarly and would not account for the asymmetrical line shapes that are observed. Finally, the simultaneous presence of a significant IR excess combined with the wavelength dependence of the line blue-shifting, as discussed by \citet{2014Natur.511..326G} and \citet{2012AJ....143...17S}, strongly favours a dust formation scenario, with dust grains condensing in a cool, dense shell between the shock fronts or in the ejecta itself. 

Further, the evolving asymmetries in the optical line profiles, first shifting towards the blue by day 112 before returning slowly towards the center over several years, is consistent with an increasing dust mass interior to the forward shock causing greater extinction by dust of emission from the receding side. During the first $\sim$250-500\,d, the strengthening blue-shift indicates that the rate of dust formation was high enough to compensate for the drop in the dust optical depth naturally caused by the expansion of the system. Whilst the blueshift persists later, its decreasing strength is likely the result of a net decrease in the dust optical depth due to expansion of the system dominating over the dust formation rate. 

In an expanding medium where lines are broadened by bulk motions, dust grains can only induce line asymmetries where the dust is interior to or co-located with the emitting source. In a Type IIn SN, red-blue asymmetries in the IWC, such as those seen in SN 2010jl, can only be caused by dust located behind the forward shock. Several dust scenarios could account for red-blue asymmetries in the IWCs of SN 2010jl, e.g. if dust formed rapidly in a CDS between the forward and reverse shocks; if pre-existing CSM dust was overrun by the blast wave and a fraction survived; and/or clumps of dust formed in the ejecta which were dense enough to survive the high X-ray luminosity of SN 2010jl \citep{2012AJ....143...17S,2015ApJ...810...32C}. Different populations of dust may account for the observed asymmetries at different times. 

Unshocked, pre-existing CSM dust cannot account for the observed line asymmetries since it is exterior to the source of the emission and therefore would attenuate the line uniformly across its width. However, the early IR excess could be produced by pre-existing dust in the unshocked CSM, warmed by the initial SN flash or by ongoing interactions. It could also be produced by dust in the ejecta or CDS heated either by radiation from the interaction or by radioactive decay. By modelling different dust populations in emission (the SED) and in extinction (the asymmetric line profiles) using two Monte Carlo radiative transfer codes, we have sought to distinguish the different dust populations and to determine their dust masses.

\section{Dust radiative transfer models of SN 2010jl}
\label{sec:models}
In this section we describe our approach to modelling the evolving SEDs and blueshifted emission line profiles of SN 2010jl using {\textsc{mocassin}} \citep[][]{2003MNRAS.340.1136E,2005MNRAS.362.1038E}, which is able to model the dust in emission, and {\textsc{damocles}} \citep[][]{2016MNRAS.456.1269B,2018MNRAS.480.4659B,2018ascl.soft07023B}, which is able to model the effects of dust in extinction on the optical and NIR line profiles, respectively. We have attempted to disentangle pre-existing circumstellar dust from dust interior to the forward shock, and to infer dust formation rates and dust masses in SN 2010jl. 

\subsection{Model geometry}
\label{sec:geometry}

\begin{figure*}
	\centering
    \includegraphics[width=0.62\textwidth,clip = true, trim = 0 10 0 0]{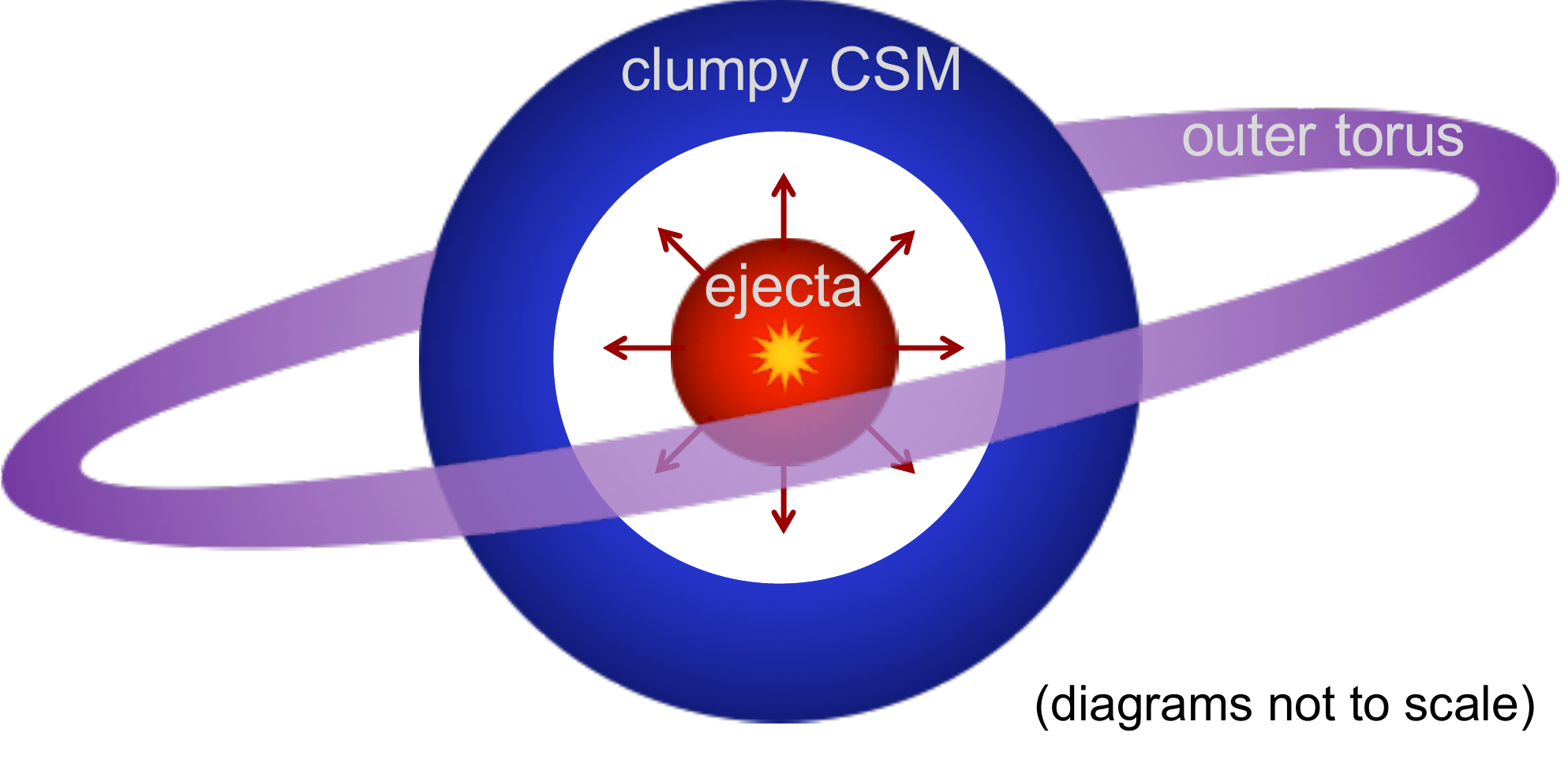}	\hspace{0.65cm}
	\includegraphics[width=0.3\textwidth,clip = true, trim = 330 260 67 20]{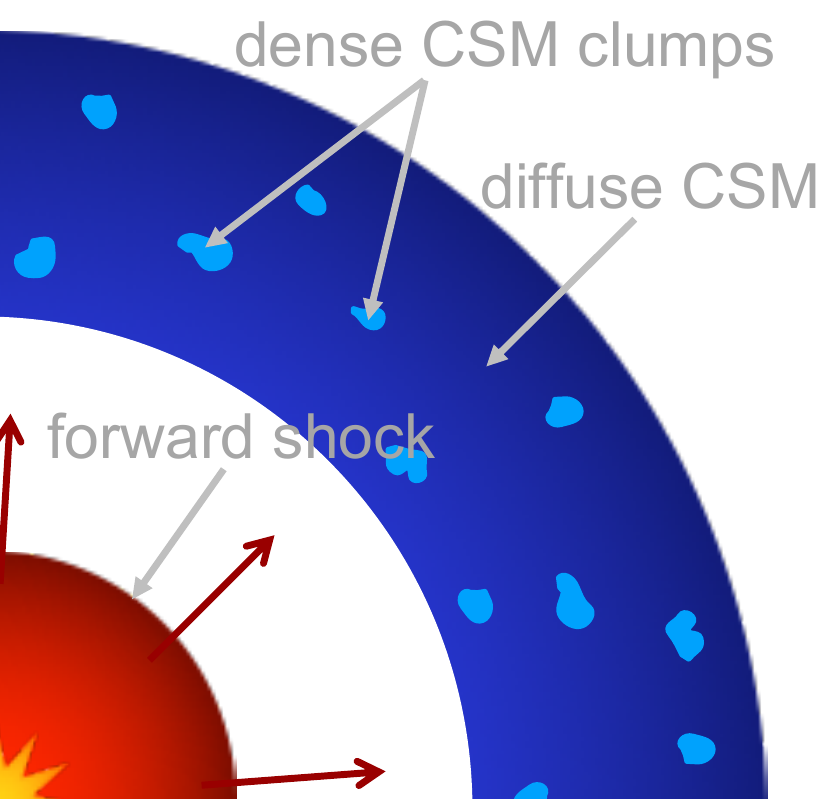} 
	\caption{A diagram of a clumpy, inner shell of CSM surrounding the expanding ejecta of SN 2010jl. Upon reaching the CSM, some of the ejecta impacts the clumps whilst some is able to travel freely through the less dense interclump medium. The dusty outer torus is responsible for the IR echo at early times. Only pre-existing dust in the outer torus survives the supernova flash. Dust forms behind the forward shock in the CSM/ejecta at later times. Similar models have been proposed for Type IIn SNe such as SN~1988Z and SN~2005ip \citep{1994MNRAS.268..173C, 2009ApJ...695.1334S}.}
	\label{Fig:schematic}
\end{figure*}

We adopted a geometry that is consistent with observations and able to reproduce both SEDs and line profiles with high quality fits. A schematic representing our adopted geometry is shown in Figure~\ref{Fig:schematic}. 

Properties of the CSM in SN 2010jl can be inferred from the presence of a strong IR excess at 91\,d post-outburst. \citet{2011AJ....142...45A} found that the IR excess at this time is likely due to echoing, pre-existing circumstellar dust. In order to reproduce the full SED at 91\,d, however, the total dust optical depth along the line of sight must be sufficiently low  to allow significant optical radiation to escape. Geometries such as a clumpy shell, or inclined lines of sight through a bipolar geometry or torus would allow for this. \citet{2011AJ....142...45A} adopted the latter geometry. Our preliminary models with {\sc mocassin} of a spherically symmetric, homologously expanding ejecta shell ruled out ejecta dust as the source of the IR excess at this epoch, supporting the conclusions of \citet{2011AJ....142...45A}. We follow their approach and investigate properties of an `outer torus' of echoing circumstellar dust at 91\,d post-outburst with the benefit of additional observational data (see Section \ref{sec:echo}). 

In addition to an outer torus of circumstellar dust, the geometry of the supernova system itself can be inferred from the line profiles. Preliminary models and careful visual inspection of the line profiles reveals the simultaneous presence of both a BWC and an IWC in certain lines from $\sim80$\,d. This is particularly clear in the He~{\sc i} 10830\AA\ line at 247\,d, and is also present in the He\,{\sc i} 20587\AA\ line, and has been discussed previously by previous authors \citep{2015ApJ...801....7B,2018MNRAS.481.3643C}. We therefore require a scenario that can produce both components simultaneously. The BWC has been frequently attributed to the effects of electron scattering in SN 2010jl and this is certainly an important effect at early times \citep{2014ApJ...797..118F}. At later times, however, the asymmetric, non-Lorentzian shape suggests a alternative origin for the BWC \citep{2014Natur.511..326G,2018MNRAS.481.3643C}. We therefore attribute the BWC to fast moving ejecta. To account for the simultaneous presence of both the BWC and the IWC, we presume that the CSM is formed of dense clumps embedded in a more diffuse inter-clump medium through which the rapidly expanding ejecta can proceed relatively unimpeded. We attribute the IWC to shocked, dense CSM clumps, as has been described by \citet{1994MNRAS.268..173C}.  The CSM clumps may become mixed with reverse-shocked ejecta over time. The early onset of interaction indicates a dense CSM close to the initial explosion. This clumped CSM, which we assume to be distributed in a shell, is therefore distinct from the more distant, echoing outer torus,  which our models show must be located at a much greater distance (see Section \ref{sec:echo}). A schematic of our adopted geometry is shown in  Figure~\ref{Fig:schematic} and further details are given below. This model is consistent with the line profile shapes and the IR echo seen at 91d. We do not preclude other geometries, and the results that we present here are one possible model that is able to self-consistently reproduce both photometric and spectroscopic observations.

We determined the initial inner radius of the CSM shell based on the appearance of the IWC in the line profiles at $\sim$80\,d. We adopted a maximum ejecta expansion velocity of 15,000\,km\,s$^{-1}$ based on the broad wings of the He~{\sc i} 10830\AA\ line at 247\,d assuming minimal contribution from electron scattering. This yielded an initial inner radius for the CSM (prior to outburst) of $R_{\rm in} \sim 1 \times 10^{16}$\,cm. The impact of the forward shock with dense CSM clumps propagates a reverse shock back into the ejecta which we adopted as the inner radius in our models. We evolved the inner radius with time based on the slowest velocities of the post-shock region (400\,km\,s$^{-1}$ inferred from preliminary line profile models with {\sc damocles}). We presume that the forward shock is able to pass through the more diffuse interclump medium without significant deceleration. We therefore adopted an outer radius of the CSM, $R_{\rm out}$, based on a forward shock velocity of 15,000\,km\,s$^{-1}$, calculated independently for each epoch. 

CSM clumps were stochastically distributed according to a density distribution $\propto r^{-2}$. Since the majority of material is concentrated towards the inner CSM radius, changes to the outer radius, such as to reduce it based on a decelerating or slower expansion velocity, do not significantly alter our results. Our CSM radius assumptions are supported by our \textsc{mocassin} models, however, which could not reproduce the SED without an extended post-shocked region stretching to large radii (see Section \ref{sec:mocassin}). Initially, a clump volume-filling factor ($f$) of 0.1 was adopted with clumps of radius $R_{\rm out}/30$. This assumption was investigated and is discussed further in Section \ref{sec:damocles}.

The geometry can be summarised as an interior shell of unshocked ejecta impacting a shell of clumpy, shocked material between the forward and reverse shocks, and surrounded by a clumpy shell of unshocked CSM. The entire system is surrounded by an outer torus of dusty CSM. Throughout this paper, we refer to the inner CSM as `CSM' and the outer torus of CSM as the `outer torus'.

The concept of the CDS becomes somewhat more complex in this scenario, with an idealised shell of rapidly cooling material seeming less likely due to the inhomogeneous nature of the CSM. Instead, we presume that the reverse shock generated by impacts between the forward shock and the dense clumps yields numerous rapidly cooling regions of dense material in which dust grains could theoretically form. For ease, we will continue to refer to such regions collectively as a `cool, dense shell' or CDS.

\subsection{DAMOCLES models of SN~2010jl}
\label{sec:damocles}

\subsubsection{Radiative transfer models of the blueshifted H\texorpdfstring{$\alpha$}{Ha}, H\texorpdfstring{$\beta$}{Hb} and \texorpdfstring{He\,{\sc i}\,5876 \AA}{He I 5876} lines of SN 2010jl}
In order to derive dust masses from observed late-time emission line profiles in CCSNe, \citet{2016MNRAS.456.1269B} developed a Monte Carlo radiative transfer code, {\sc{damocles}}, which models the scattering and absorption of line photons by dust in supernovae. {\sc{damocles}} allows for arbitrary dust density, emissivity and velocity distributions to be specified and any number of dust species and grain size distributions to be investigated. 

A range of models treating multiple lines in the spectrum (H$\alpha$, H$\beta$, and He\,{\sc i} 5876 \AA) simultaneously were developed for three epochs: 526, 915, and 1286 days. These epochs were selected for their coverage of the period of interest and their quality. The lines modelled represented the strongest lines in the spectra and were not significantly contaminated by other nearby lines. By fitting these observed line profiles at a range of epochs, we were able to estimate the mass of dust present in  the CDS in SN2010jl and its evolution over time.  

\begin{table*}[!htbp]
	\caption{Best-fitting parameters for the two velocity component clumpy {\sc{damocles}} models. The parameters are defined as follows: $v_{\rm max,\,IWC}$ and $v_{\rm min,\,IWC}$ are the maximum and minimum velocities for the IWC, respectively. $\alpha$ is the radius-independent gradient of the velocity distribution for the IWC, see Equation \ref{eqn:vel}. $\beta$ is the gradient of velocity distribution for the BWC which has a maximum velocity of 15,000 km s$^{-1}$ and a minimum velocity equal to the IWC maximum velocity ($v_{\rm max,\,IWC}$) for continuity (see Equation \ref{eqn:vel}). $M_{\rm dust}$ is the total mass of the dust and $R_{\rm out}$ and $R_{\rm in}$ are the outer and inner radii of the post-shock region, respectively. }
	\def\arraystretch{1.15}
	\centering
	\begin{tabular}{cccccccccc}
		\\
		\hline
		line & epoch & $v_{\rm max,\,IWC}$ & $v_{\rm min,\,IWC}$ & $\alpha$ &  $\beta$ & a &M$_{\rm dust}$  & R$_{\rm in}$ & R$_{\rm out}$\\
		& d & km s$^{-1}$ & km s$^{-1}$ & & & \micron & 10$^{-3}$ M$_{\sun}$ & 10$^{16}$ cm & 10$^{16}$ cm \\
		\hline
		H$\beta$ & 526 & 2300 & 650 & 0.1 &  -4.5 & 0.1&0.25  & 1.19 & 6.82\\
		H$\alpha$ & 526 & 2300 & 650 & 0.1 & -4.5 & 0.1&0.25  & 1.19 & 6.82\\
		\\
		H$\beta$ & 915 & 1800 & 550 & 0.3  & -4.0 & 0.2&2  & 1.33 & 11.9\\
		H$\alpha$ & 915 & 1800 & 550 & 0.3  & -4.0 &0.2& 2  & 1.33 & 11.9\\
		\\
		H$\beta$ & 1286 & 1500 & 500 & 0.6  & -4.0 & 0.2&5  & 1.46 & 16.7\\
		 H$\alpha $ & 1286 & 1500 & 500 & 0.6  & -4.0 &0.2& 5  & 1.46 & 16.7\\
		He\,{\sc i}\,$\lambda 5876$ & 1286 & 1500 & 200 & 0.1 & -4.0 &0.2 &5  & 1.46 & 16.7 \\
		\hline
		\\
	\end{tabular}
	\centering
	\label{tb_damocles}
\end{table*}

Line photons were emitted from a shell with outer radius $R_{\rm out}$ and inner radius $R_{\rm in}$ representing the positions of the forward and reverse shocks, respectively, as discussed in Section \ref{sec:geometry}. An emissivity law $\propto r^{-4}$ was applied, appropriate for recombination in a medium with an inverse-square gas density law, as might be expected for a CSM produced by mass loss at a constant rate. As long as material was predominantly concentrated in the inner regions of the shell, variations in the density distribution of the emitting material did not significantly affect the results. 

Due to the complex nature of dense clumps accelerated by the forward shock, with reverse-shocked ejecta mixing with these CSM clumps due to instabilities caused by the impact, and fast moving ejecta moving through more diffuse regions between clumps, the typical $v\propto r$ velocity distribution is likely not appropriate in this post-shock region. We therefore parameterized the velocity distribution as a power-law velocity $p(v) \propto v^{\alpha}$ independent of radius, with the velocity of an emitting particle sampled from this distribution for every propagated packet. We required that each emitting species followed the same velocity law, but allowed different velocity distributions for hydrogen and helium. The same minimum and maximum velocities were imposed for all emitting species.

The steepness of the velocity law, and the minimum and maximum velocities were varied for all three lines simultaneously until good fits to the blue side of the IWC were acquired. 
Dust was then introduced to the models. We used amorphous carbon grains with optical constants given by the BE sample from \citet{1996MNRAS.282.1321Z}. The lack of observable emission features around 10~$\mu$m suggests that the fraction of silicate dust in SN 2010jl is limited and so we considered only amorphous carbon grains for these models \citep{2018ApJ...859...66S}. We discuss this assumption further in Section \ref{sec:discussion}. 

In order to simultaneously fit H$\alpha$, H$\beta$, and He\,{\sc i}\,5876 \AA, the grain radius of the dust was varied to account for the wavelength dependence of the dust optical depth to which each line was exposed. We found that the dust optical depth to which the H$\alpha$ and H$\beta$ lines were exposed was very similar, placing strong constraints on the grain radius. A single grain radius was used for these models and, under this assumption, a flat wavelength dependence could only be achieved using grains of at least $a\geq 0.1\,$\micron. Smaller grains have a steeper wavelength dependence that could not  simultaneously fit the lines. For smooth models, larger grains would also be inappropriate since they would be too scattering and produce a significant red scattering wing inconsistent with the line profiles. However, clumping suppresses this effect and thus larger grains may be possible in our line profile models. Further constraints to the grain radius can be derived from the {\sc Mocassin} models, which restrict the grain radius to $a\leq0.3$~\micron\ in all cases. By optimising both the line profiles fits and the SED fits using a single grain size, we infer grain radii of 0.1~\micron\ at 526\,d and 0.2~\micron\ at 915\,d and 1286\,d. It should be noted that distributions of grain radii might alter this inference. 

Our conclusions are not necessarily in conflict with suggestions by other authors that there is evidence of a strong wavelength dependence in line extinction since their analyses are based on a wider wavelength range than we have modelled here and they also inferred only small variations in extinction  over the wavelength range we consider here \citep{2012AJ....143...17S,2013MNRAS.435.1520M,2014Natur.511..326G}.  
\begin{figure*}
	\centering
	\includegraphics[width=0.36\textwidth,clip=true,trim=7 4 7 7 ]{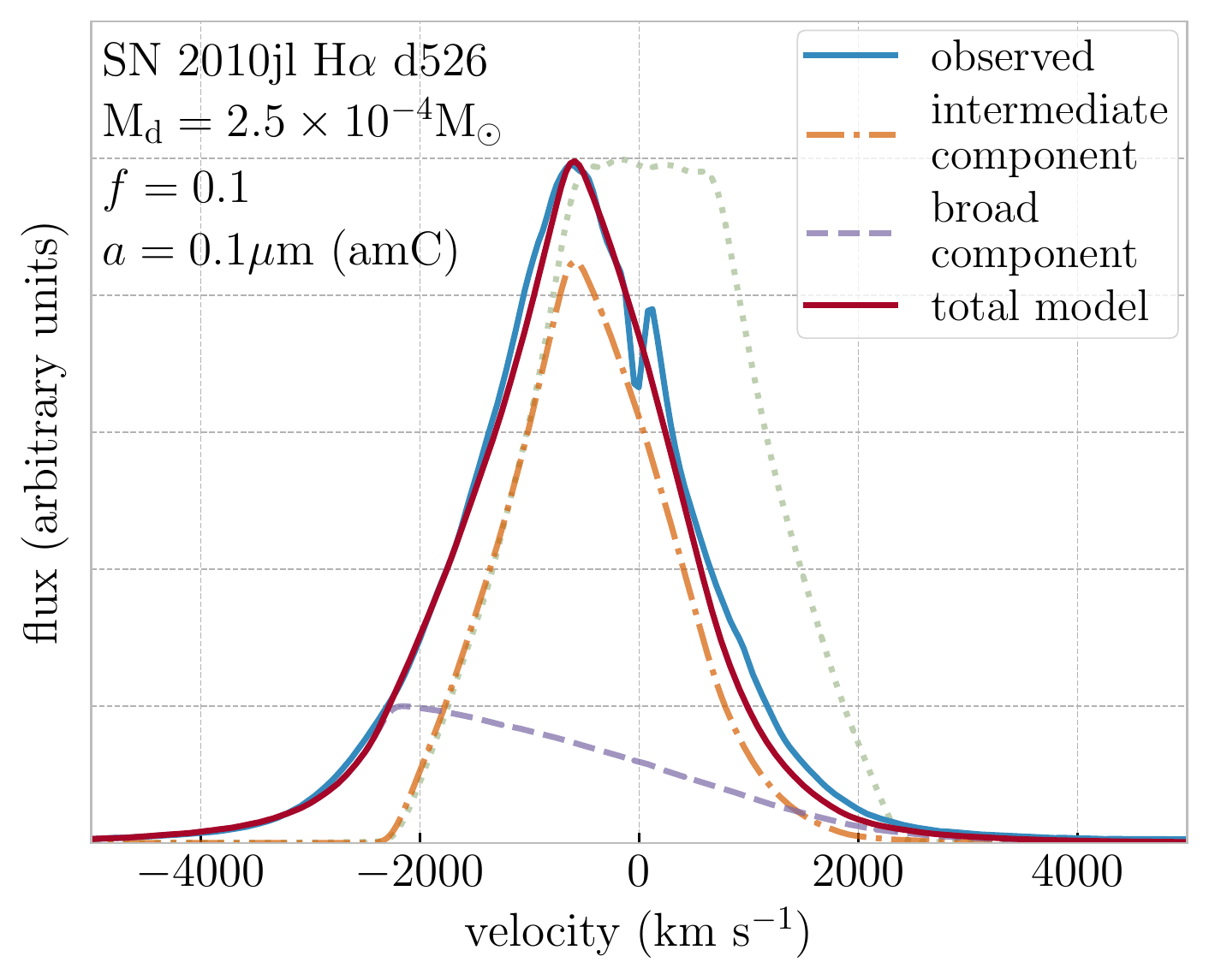}  \hspace{2cm}
	\includegraphics[width=0.36\textwidth,clip=true,trim=7 4 7 7 ]{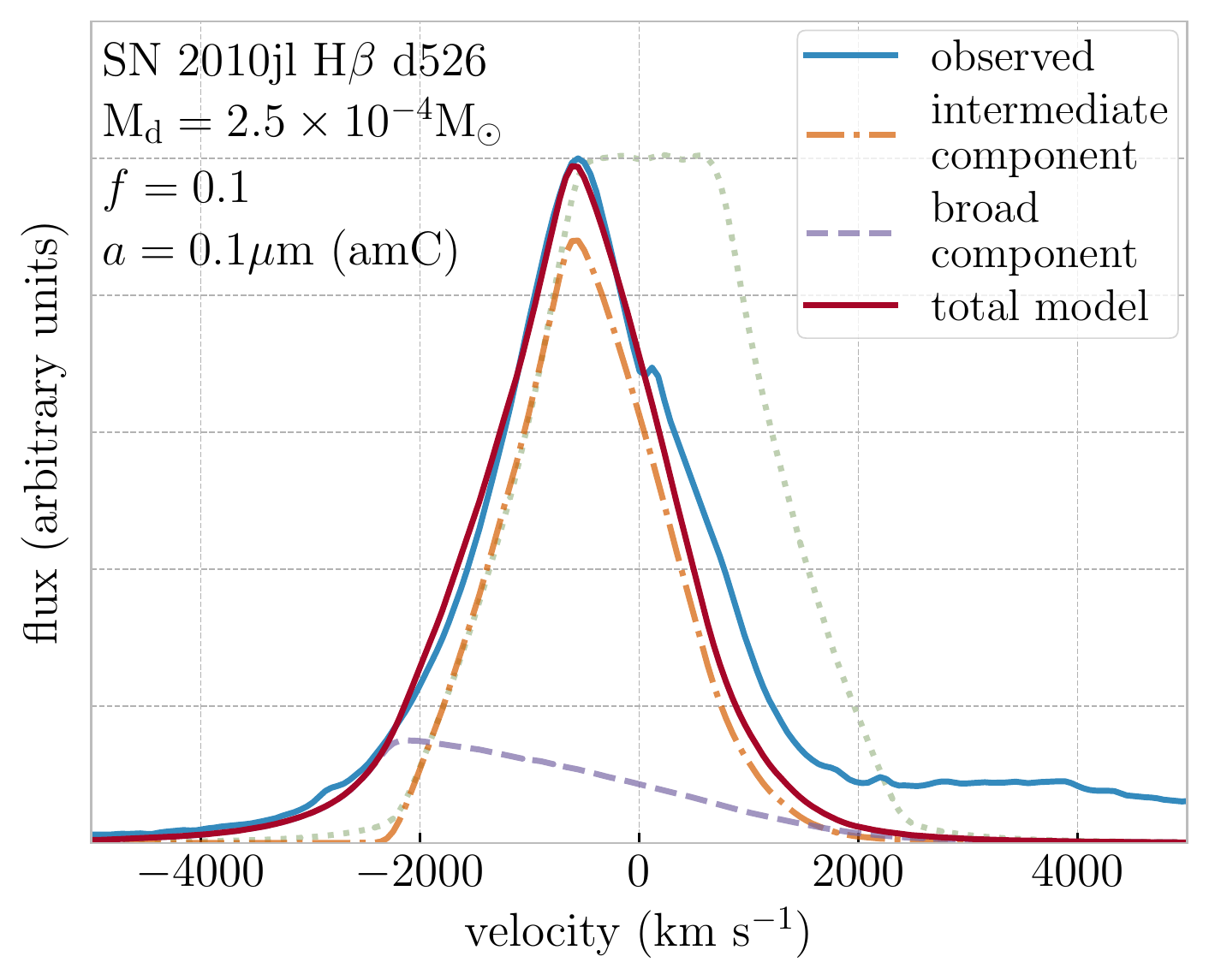}
	\includegraphics[width=0.36\textwidth,clip=true,trim=7 4 7 7 ]{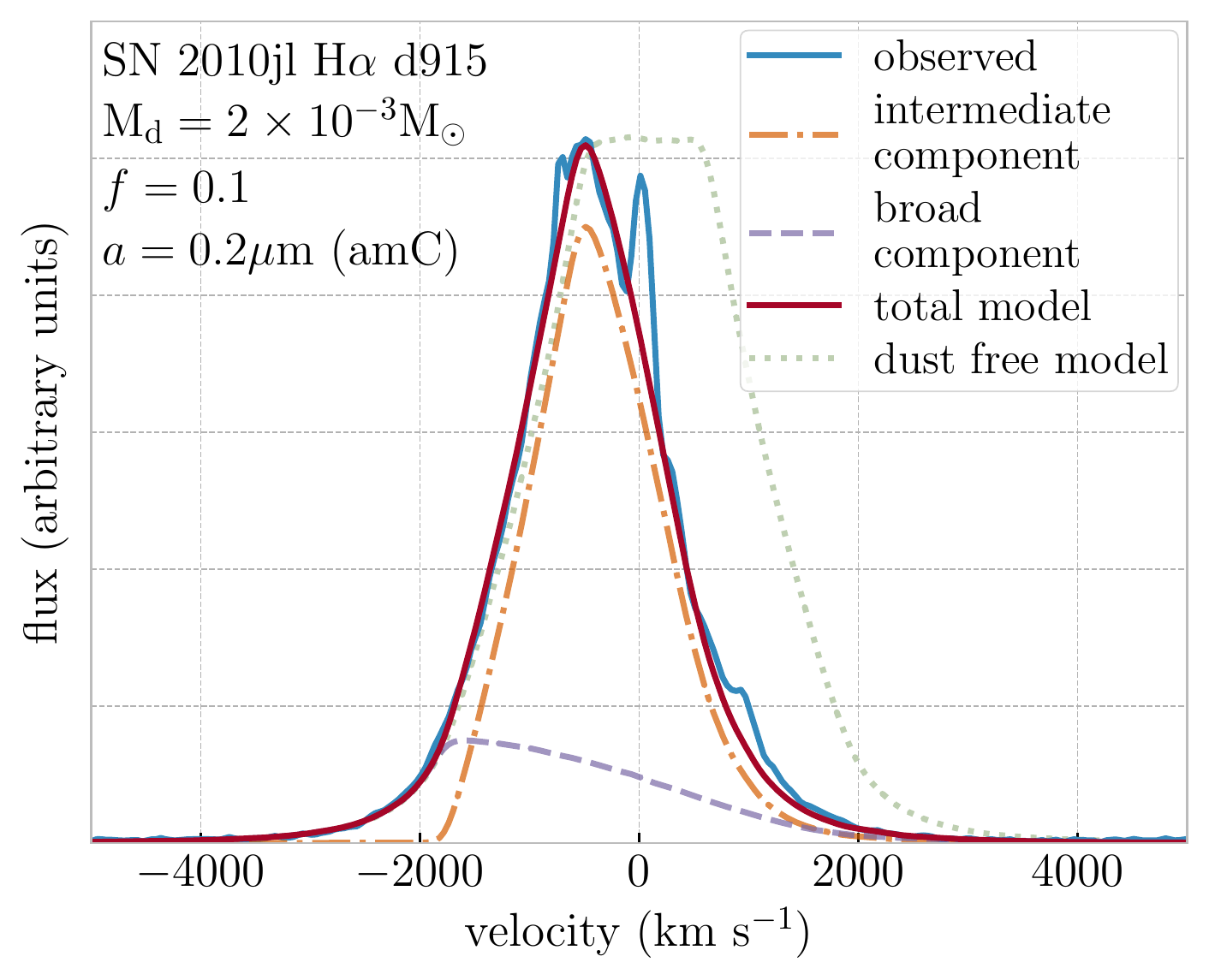} \hspace{2cm}
	\includegraphics[width=0.36\textwidth,clip=true,trim=7 4 7 7 ]{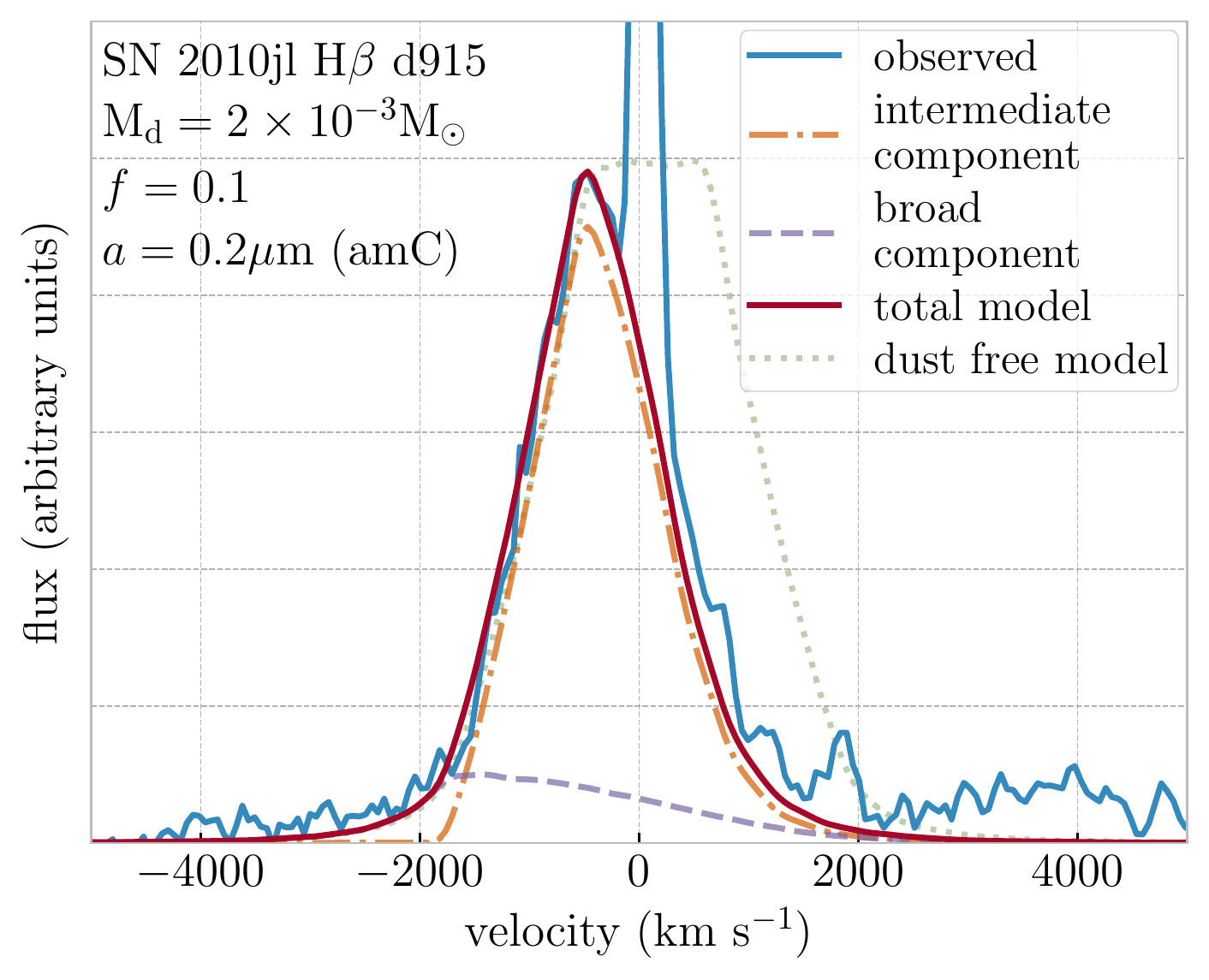} 
	\includegraphics[width=0.36\textwidth,clip=true,trim=7 4 7 7 ]{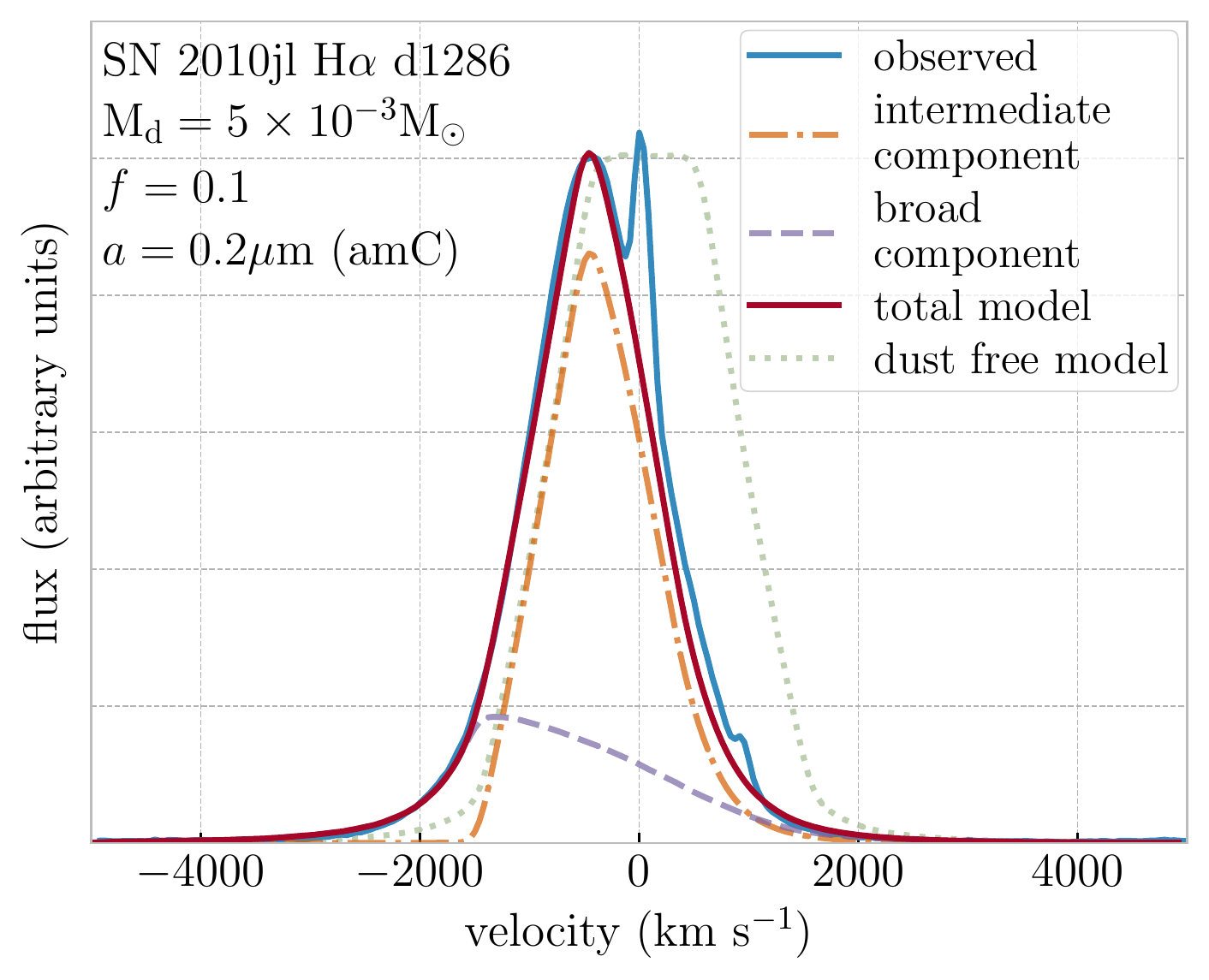} \hspace{2cm}
	\includegraphics[width=0.36\textwidth,clip=true,trim=7 4 7 7 ]{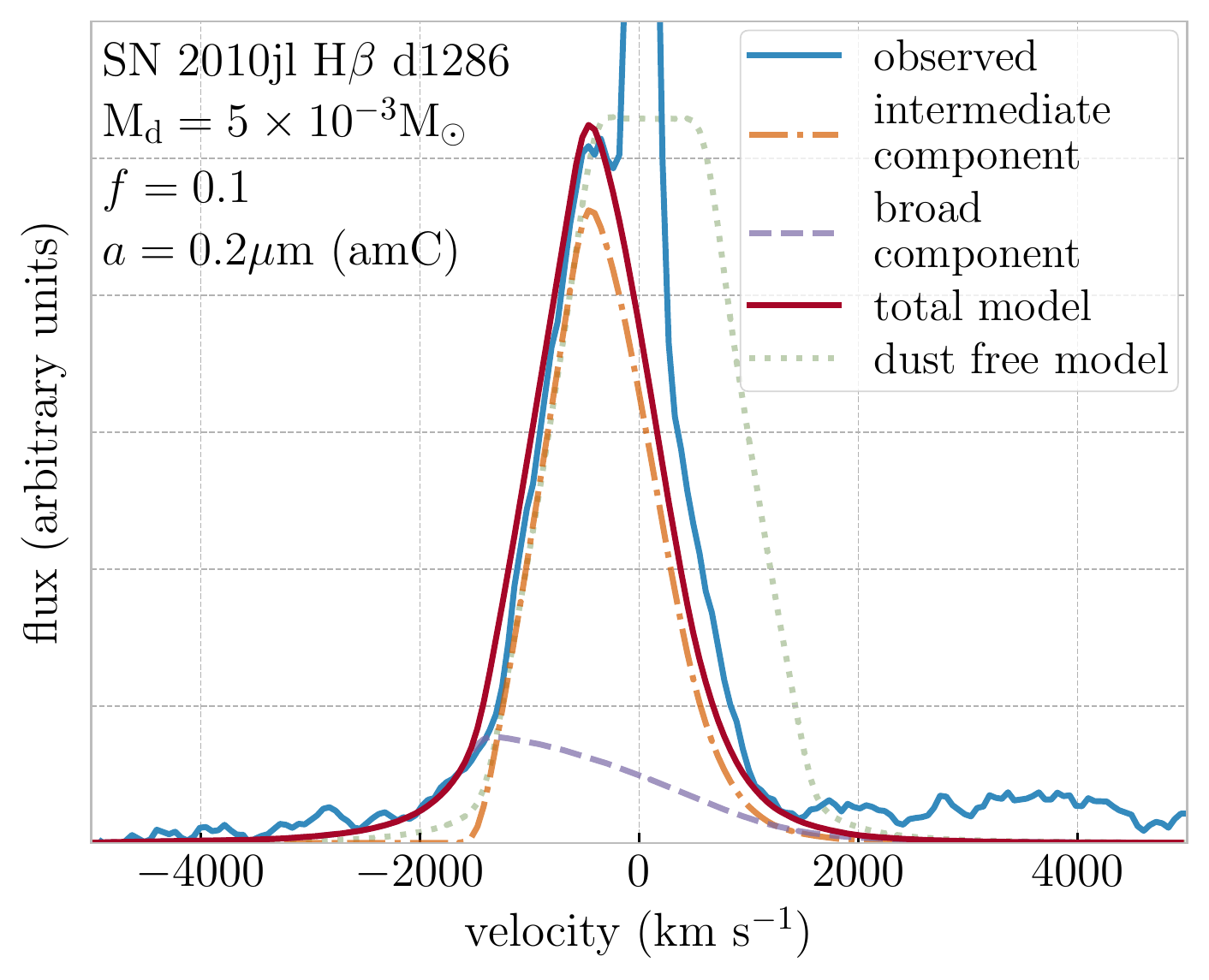}
	\includegraphics[width=0.36\textwidth,clip=true,trim=7 4 7 7 ]{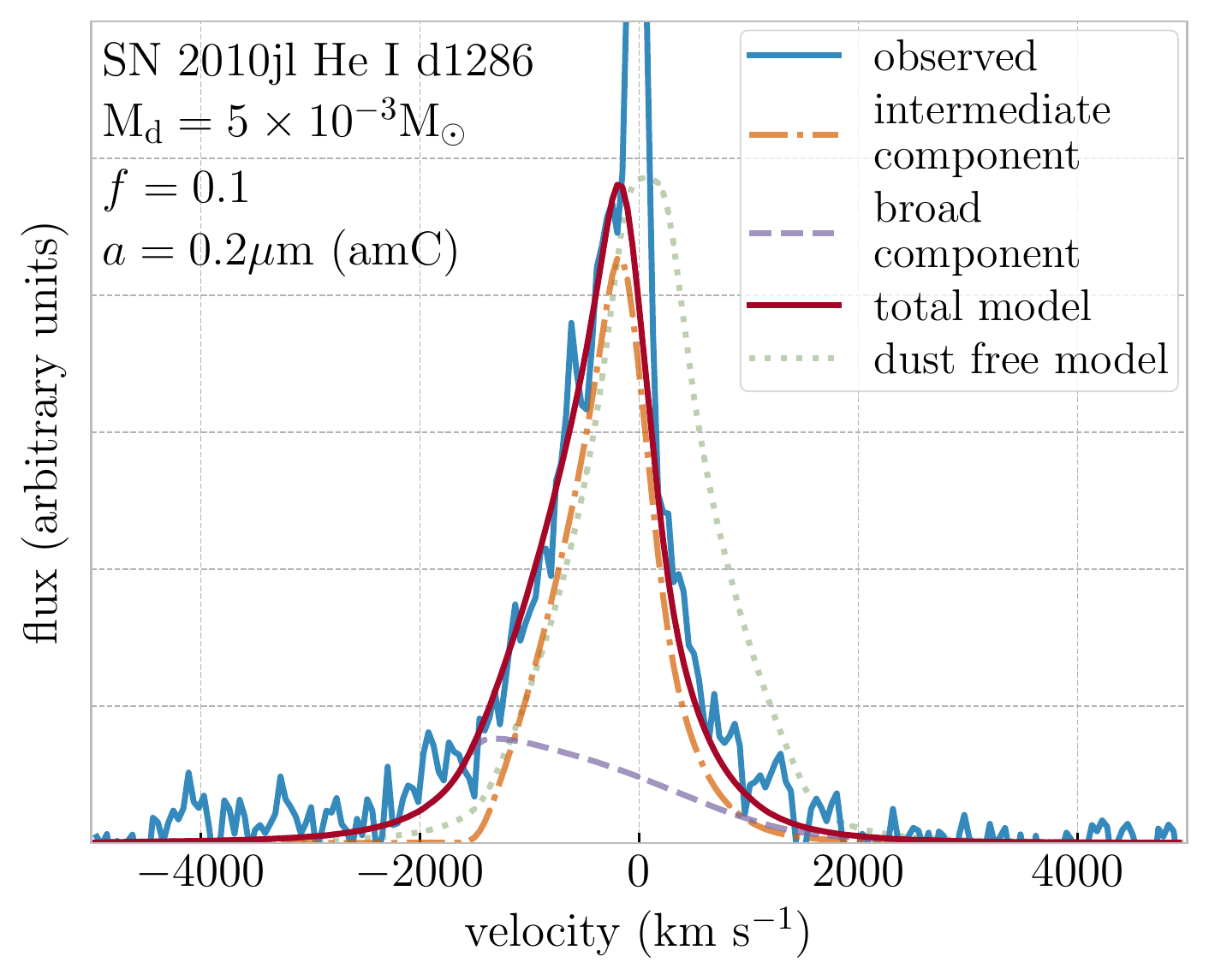} 
	\caption{ {\sc{damocles}} models of the IWC and BWC of SN 2010jl's strongest optical emission lines, presented in velocity space, for 3 epochs: 526\,d, 915\,d, and 1286\,d. $M_{\rm dust}$ is the total mass of dust, $a$ is the single size grain radius for amorphous carbon (amC) grains, and $f$ is the clump volume filling factor set at $f=0.1$. The narrow emission lines in the observed line profiles (\textit{blue solid lines}) indicate zero-velocity.  {\sc{damocles}} models for the IWCs (\textit{orange dash-dot lines}) and the BWCs (\textit{purple dashed lines}) are combined to produce the final models (\textit{solid red lines}). Scaled, intrinsic dust free models are also shown {\it (green dotted lines)}.}
	\label{Fig:twodamocles}
\end{figure*}

The resulting profile fits reproduced the dominant IWC component well, but could not fit the wings of the profile. We therefore considered a modified velocity law, introducing an additional broad component representing the BWC produced by the fast-moving ejecta. Using a similar prescription to the IWC models, we added an additional radius-independent, power-law velocity component at higher velocities such that the velocity distribution was
\begin{equation}
\label{eqn:vel}
p(v) \propto \begin{cases}
v^{\alpha}, & \text{$v_{\rm min,\,IWC}<v<v_{\rm  max,\,IWC}$}\\
v^{\beta}, & \text{$v_{\rm max,\,IWC}<v<15,000{\rm \, km \, s}^{-1}$},
\end{cases}
\end{equation}
\noindent where we impose that the minimum velocity of the broad component is equal to $v_{\rm max,\,IWC}$ for continuity. We did not change the dust distribution from previous models fixing $f=0.1$, which we also used for our clumpy {\sc{mocassin}} models. The broad line emission, from the ejecta expanding between the clumps in the CSM, was emitted from between the same inner and outer radii as the IWC. The dust distribution was left unchanged. The introduction of this additional velocity component required only minor changes to the previously best-fitting parameters inferred from just a single IWC. All lines can be simultaneously fitted using the same dust distribution.

Best fits in all cases were obtained through chi-square minimization to the continuum-subtracted emission line profiles. We estimate the uncertainties in the dust masses to be $\sim$30\%. The uncertainties are based on a 10\% variation in $\chi^{2}$ when varying only the dust mass and keeping all other parameters fixed. The He~{\sc i} 5876 \AA~line was noisy at the earlier epochs and so was only considered for verification purposes except at the final epoch at  1286\,d. 

The final fits to the emission line profiles are shown in Figure~\ref{Fig:twodamocles}, and the parameters and dust masses are listed in Table \ref{tb_damocles}. These models trace dust located in the CDS in the post-shock region. The presence of dust in this region of an interacting supernova can theoretically be explained in two ways; either new dust grains have formed from heavy elements in the shocked CSM or shocked ejecta, or pre-existing CSM dust  survived the passage of the blast wave. Our models of the light echo at 91\,d (see Section \ref{sec:echo}) preclude the presence of pre-existing dust at radii close to the location of the forward shock ($\sim0.7 - 2 \times 10^{17}$cm) and we therefore deduce that the dust in the CDS of SN 2010jl is newly-formed dust that has condensed in rapidly cooling regions inside dense clumps.

\subsubsection{Optically thick dust clumps}

One possible limitation of using line profiles to infer dust masses is the possibility that the dust clumps might have high optical depths, obscuring the presence of large masses of dust in their interior \citep{2015ApJ...810...75D,2019ApJ...871L..33D}. These clumps could be transparent in the IR, and thus visible in emission but not in extinction. Further, highly optically thick dust washes out any wavelength dependent absorption and makes it difficult to determine properties of the dust grains. We investigated the possibility of optically thick clumps in our models.

\begin{figure}
    \centering
    \includegraphics[width=0.49\textwidth,clip=true,trim = 23 6 60 35]{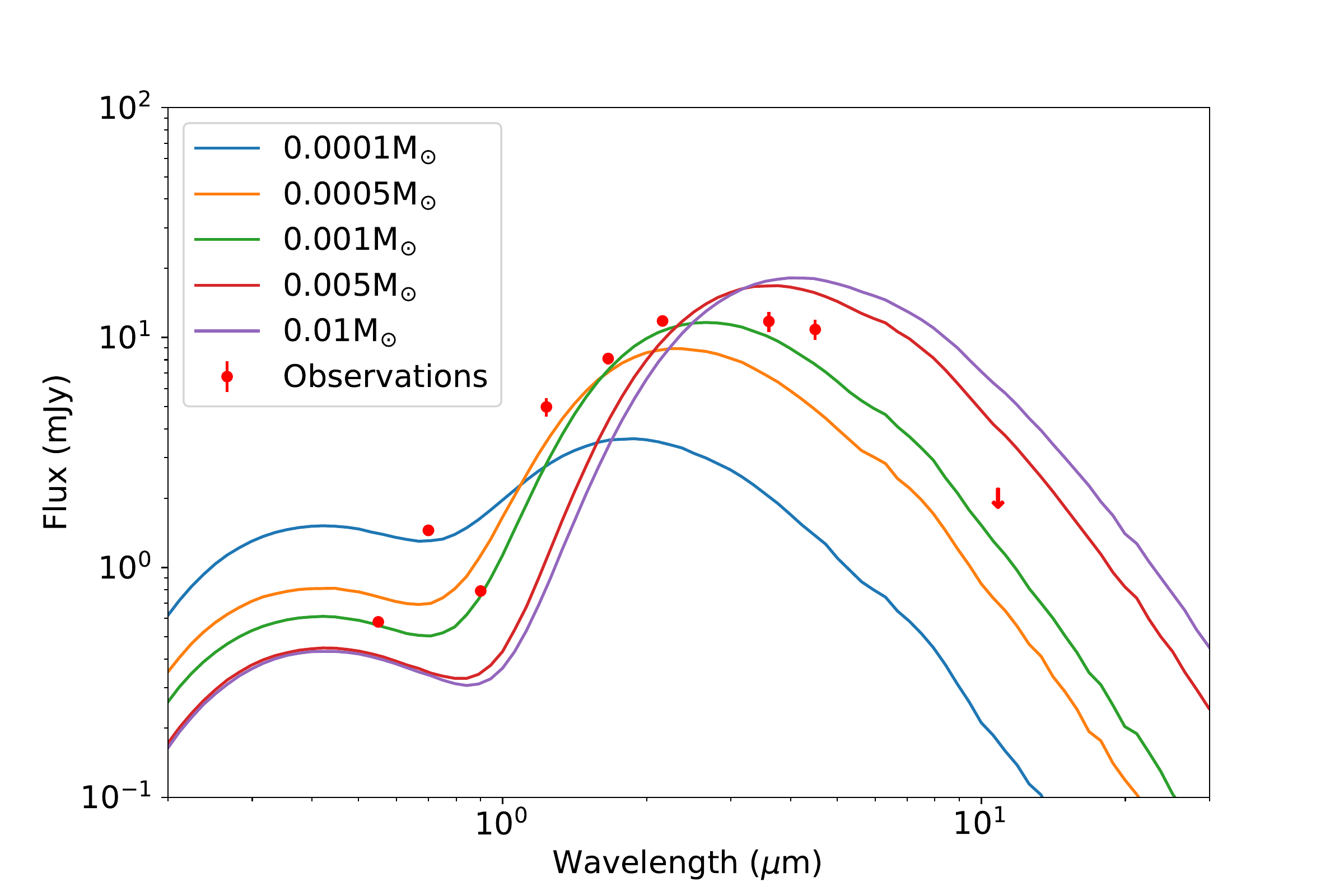}
    \caption{{\sc mocassin} SEDs for a clump filling factor of $f$=0.03 and observations at 464\,d. Emission line profiles are not sensitive to the dust mass for this particular filling factor, but the SED remains so.}
    \label{fig:mocassinffsensitivity}
\end{figure}

Our initial models adopted a volume-filling factor ($f$) of 0.1 with clumps of radius $R_{\rm out}/30$. We varied the filling factor over a range of values $<$0.3. Filling factors with $f <$ 0.03 only cover a small area of the supernovae and so do not obscure enough optical emission to reproduce the asymmetries seen in the line profiles. For $f >$ 0.03, the line profiles can be reproduced in all cases and, crucially, are still sensitive to the mass of dust in the clumps. Large masses of dust cannot be hidden in clumps when the filling factor is $>0.03$ and so the dust mass can still be constrained. The dust mass that is required to fit the line profiles does vary slightly as the filling factor is varied between $0.03<f<0.3$ by around $\sim 30$\%.
 We found only one, very specific case where large masses of dust could be concealed inside optically thick clumps. This was when the filling factor was exactly $f$ = 0.03. This specific filling factor results in a dust distribution that covers enough of the supernova to attenuate sufficient optical radiation to reproduce the profiles, but also results in clumps that are sufficiently optically thick that they can conceal large (theoretically infinite) amounts of dust. In this case, neither the grain radius nor dust mass can be constrained. However, {\sc Mocassin} models of the SED using a volume filling factor of $f=0.03$ are not consistent with large dust masses and are still able to constrain the dust mass, removing this degeneracy (Figure~\ref{fig:mocassinffsensitivity}). Our final results for the two-component clumped line profile models with $f$ = 0.1 are given in Table \ref{tb_damocles} and fits can be seen in Figure~\ref{Fig:twodamocles}.

\begin{figure}
    \centering
    \includegraphics[width=0.49\textwidth,clip=true,trim = 23 6 60 35]{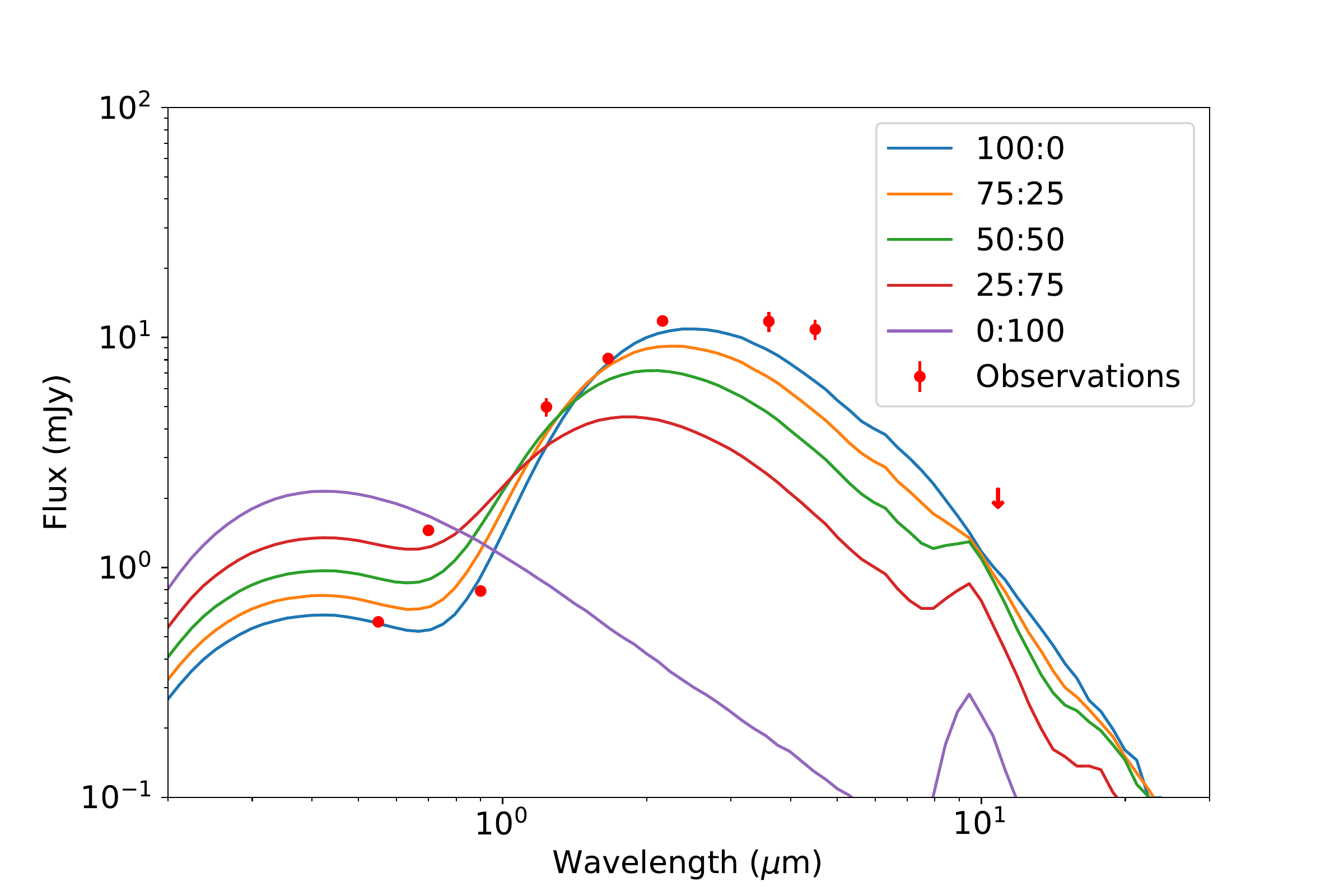}
    \caption{Predicted SEDs for 7$\times$10$^{-4}$M$_\odot$ of dust with composition indicated in the key as the amorphous carbon:silicates ratio. Observations are at 464\,d. Non-zero silicate fractions result in too much optical emission escaping and not enough infrared emission to fit the observed excess.}
    \label{fig:silicatesruledout}
\end{figure}

\begin{figure*}[!ht]
	\includegraphics[width = \textwidth,clip = true, trim=70 20 80 70]{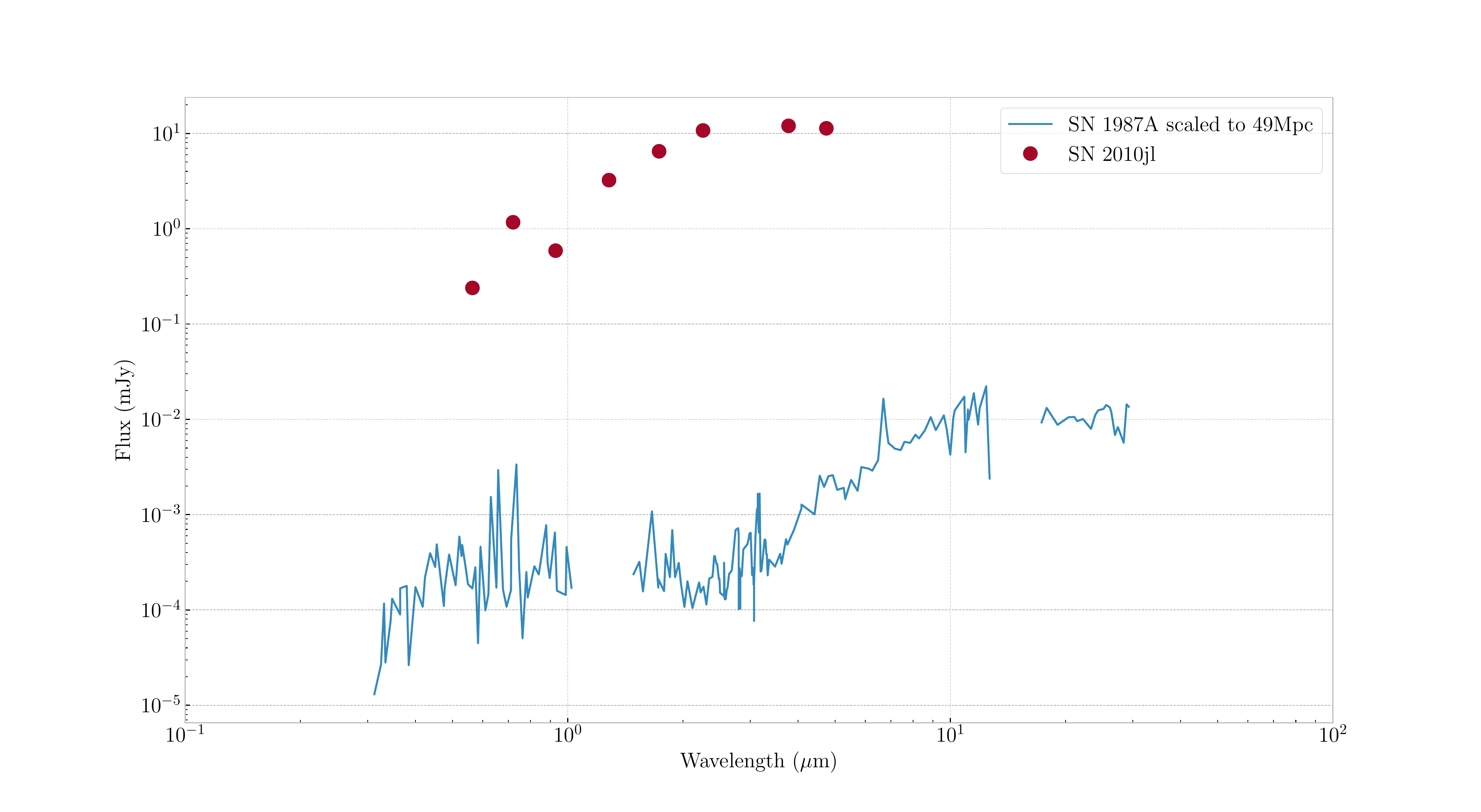}
	\caption{The SEDs of SN 1987A \citep{1993ApJS...88..477W, 2015MNRAS.446.2089W} and SN 2010jl on days 615 and 621, respectively. The flux densities from SN 1987A are scaled for the distance of SN 2010jl. While SN 2010jl was an X-ray and optically luminous SN, it also has a very large IR excess. Its dust luminosity is too large to only be dust warmed by radioactive decay.}
	\label{Fig:87Av10jl}
\end{figure*}

\subsection{MOCASSIN models for the SEDs of SN2010jl}
\label{sec:mocassin}

We used the 3D Monte Carlo radiative transfer code {\sc mocassin} \citep{2003MNRAS.340.1136E,2005MNRAS.362.1038E} to investigate the properties and mass of dust in SN 2010jl. {\sc mocassin} self-consistently calculates dust temperatures and emission based on any input spectrum, emissivity distribution and dust grid \citep{2003MNRAS.340.1136E}. For our models of SN2010jl, the grid was populated with dust with a specified grain size, number density, and composition. The model parameters were then varied to achieve an output SED that matched the observed photometry. The R-band was excluded from the fits since it is contaminated by the strong H$\alpha$ emission. It is noteworthy that the models can only put constraints on warm dust since there are no observations longward of 4.5 \micron. Colder dust may be present and emitting at longer wavelengths.

\subsubsection{Amorphous carbon dust grains}
As per our \textsc{damocles} models, we adopt a pure amorphous carbon dust composition using optical constants from the BE sample of \citet{1996MNRAS.282.1321Z}. There have been non-detections of SN 2010jl at 10.7~$\mu$m with VLT/VISIR on day 519 (see Section \ref{sec:obs}) and at 11.1~$\mu$m with {\it SOFIA}/FORCAST on day 1304 \citep{2015ApJ...808L..22W}. These non-detections provided upper limits of 2 mJy and 4.2~mJy, respectively, on the brightness of any 9.7 \micron\ silicate feature present in SN 2010jl. Models of optically thin silicate dust emission by \citet{2015ApJ...808L..22W} predicted a flux at 11\micron\ significantly exceeding  these upper limits.  Our own {\sc Mocassin} models of silicate dust in dense clumps confirm that silicate dust gives very poor fits to the SED. Figure~\ref{fig:silicatesruledout} shows the predicted emission from 7$\times$10$^{-4}$M$_\odot$ of dust, composed of pure silicates, pure amorphous carbon, and three mixed compositions. These SEDs show that an amorphous carbon fraction close to 1 is required to provide clumps which are opaque at optical wavelengths and emit efficiently enough to reproduce the observed infrared excess. Even significantly larger masses of silicate dust ($>0.01$\,M$_{\sun}$) are unable to sufficiently attenuate the optical. 
It is possible that silicates make up a non-zero fraction of the grain composition, but we infer that carbon grains dominate the emission and therefore use 100\% amorphous carbon grains in our models. We investigated a range of single grain sizes $\geq$0.1 \micron\ based on our {\sc {damocles}} models. We adopted the grain size which yielded optimal fits for both the SED and line profiles at each epoch.

\subsubsection{How is the dust heated?}
There are several mechanisms that could be heating the dust, depending on its location. Radioactive decay would only affect the ejecta dust, and we do not account for this explicitly in our models. Even if our models did account for it, the very high IR luminosity means that ejecta dust heated by radioactive decay cannot be significantly contributing. To emphasize the size of the IR excess, a comparison of the SEDs for SN 1987A on day 615 and SN 2010jl on day 621 are presented in Figure~\ref{Fig:87Av10jl}. Radioactive heating depends on the nickel mass, and for SN 1987A that is about 0.1 M$_{\sun}$ \citep{2014ApJ...792...10S}. In Figure~\ref{Fig:87Av10jl}, the SEDs are at the same epoch and scaled to the same distance and there is about four orders of magnitude more dust luminosity from SN 2010jl. \citet{2019PASP..131e4204O} preformed a similar exercise by showing that 1 M$_{\sun}$ of $^{56}$Ni was not enough to reproduce the light curve.

A far more important contributor to heating the dust is the interaction between the ejecta and the CSM. The high energy photons, emitted by the interaction, will heat the surrounding dust, new and pre-existing. Our models include a simplified version of this interaction region in {\sc{mocassin}} by using a diffuse light source that is co-located with the dust, simulating the dust being heated by the gas which is itself heated by the energetic interaction. The variables for this source of photons are temperature, luminosity, radius, and number of photons. Adopting the same radii as the {\sc Damocles} models, we include an internal, clumpy shell representing the post-shocked region as well as an outer torus of flash-heated dust.  

A third heating source is the supernova flash, seen as a thermal echo in the outer torus. \citet{2011AJ....142...45A} took the IR excess at 91\,d together with the very small line of sight extinction to imply an asymmetric CSM, with no dust along the line of sight. They adopted an inclined torus of CSM. The models described in that paper predict the evolution of the IR flux as the flash propagates through the torus. 

\subsubsection{The evolution of the thermal echo}
\label{sec:echo}
The first epoch we modeled with { \sc{mocassin}} was 91\,d. Building on the work of \citet{2011AJ....142...45A}, we were able to fit the IR excess at this epoch with an IR echo from the supernova flash illuminating an outer torus of CSM (see Figure \ref{Fig:schematic}). The 4.5$\mu$m flux is higher than the 3.6$\mu$m flux at this epoch, requiring the dust to be relatively cool and thus ruling out ejecta dust.
We used the following flash source parameters: a central point source with a blackbody continuum of temperature of 15,000 K and a luminosity of 8$\times10^{9}$ L$_{\sun}$. For the echo model, we began with the same parameters as \citet{2011AJ....142...45A}, who used the combination of negligible line-of-sight optical depth and infrared excess to propose that the CSM was in a torus, with a radius of 1.2~ly and a tube radius of 0.5~ly, inclined at 60$^{\circ}$ to the plane of the sky. The additional data available since then provide further constraints, which we used to refine our model. In particular, the non-detection of SN2010jl at 10.7~$\mu$m on day 519 means that the CSM must have a smaller outer radius than initially proposed by \citet{2011AJ....142...45A}. We find that a radius of 1.0~ly and a tube radius of 0.3~ly fits the day 91 SED while not violating the constraints at day 519. This revised geometry has its inner edge at the same distance as the earlier model. The inner edge of the torus is significantly beyond the location of the forward shock at all epoch considered in this paper. Any pre-existing dust located at radii inside the torus would be too hot to be consistent with the SED at 91\,d. We therefore infer that there is no pre-existing CSM dust interior to this torus. We find that a total dust mass of 0.015~M$_\odot$ is required to fit the SED. We assumed that the outer torus dust mass estimate at 91\,d is the total mass of dust in the torus, and that it does not change from this onwards.  

At later epochs, the total SED therefore consists of the echo from the outer torus plus emission from newly-formed dust in the CDS and/or ejecta located interior to the forward shock.  The echo models were calculated based on a flash of 100 days duration and integrating the emission from grid cells that lie between the two ellipsoids that corresponded to the beginning and end of the flash. The echo accounts for the entirety of the IR excess at 91\,d, makes a small contribution at 450-550\,d and is no longer present by 996\,d (see Figure \ref{Fig:mocassin}).

\subsubsection{Radiative transfer models of the SEDs of SN 2010jl}
\begin{table*}[!ht]
	\def\arraystretch{1.15}
	\caption{Best-fitting parameters for the  {\sc{Mocassin}} models. The parameters are defined as follows: $T_{\rm source}$ and $L_{\rm source}$ are the temperature and luminosity of the illuminating blackbody respectively,  $f$ is the clump volume filling factor, $a$ is the single grain radius, and $R_{\rm out}$ and $R_{\rm in}$ are the outer and inner radii of the post-shock region, respectively. Note that the parameters given for 91\,d are solely for echoing dust in the outer torus, the inner and outer edges of which are specified by R$_{\rm in}$ and R$_{\rm out}$ and marked with a $^\dagger$.}
	\centering
	\begin{tabular}{cccccccccc}
		\\
		\hline
		Age & T$_{source}$ & L$_{source}$ & $f$ & a & New dust mass & Echoing dust mass & R$_{\rm in}$ & R$_{\rm out}$\\
		days & 1000 K & 10$^{9}$ L$_{\sun}$ & &\micron& M$_{\sun}$ &M$_{\sun}$ & 10$^{16}$cm & 10$^{16}$cm \\
		\hline
		91 & 17.5 & 7.97 & 1.0 & 0.02 &-& 0.015 & 66.2$^\dagger$ & 123$^\dagger$ \\
		\\
		464 & 12 & 1.8 & 0.1 &0.1& 7\,$\times 10^{-4}$& 0.015 & 1.17 & 6.0 \\
		526 & 12 & 1.8 & 0.1 &0.1& 7\,$\times 10^{-4}$& 0.015 & 1.19 & 6.8 \\
		996 & 12 & 1.2 & 0.1 &0.2& 3\,$\times 10^{-3}$& 0.015 & 1.36 & 12.9 \\
		1367 &30 & 0.4 & 0.1 &0.2& 1\,$\times 10^{-2}$& 0.015 & 1.48 & 17.7 \\
		\hline
		\\
	\end{tabular}
	\centering
	\label{tb_mocassin}
\end{table*}

\begin{figure*}[!ht]
    \centering
	\includegraphics[width=0.49\textwidth,clip=true,trim = 23 6 60 35]{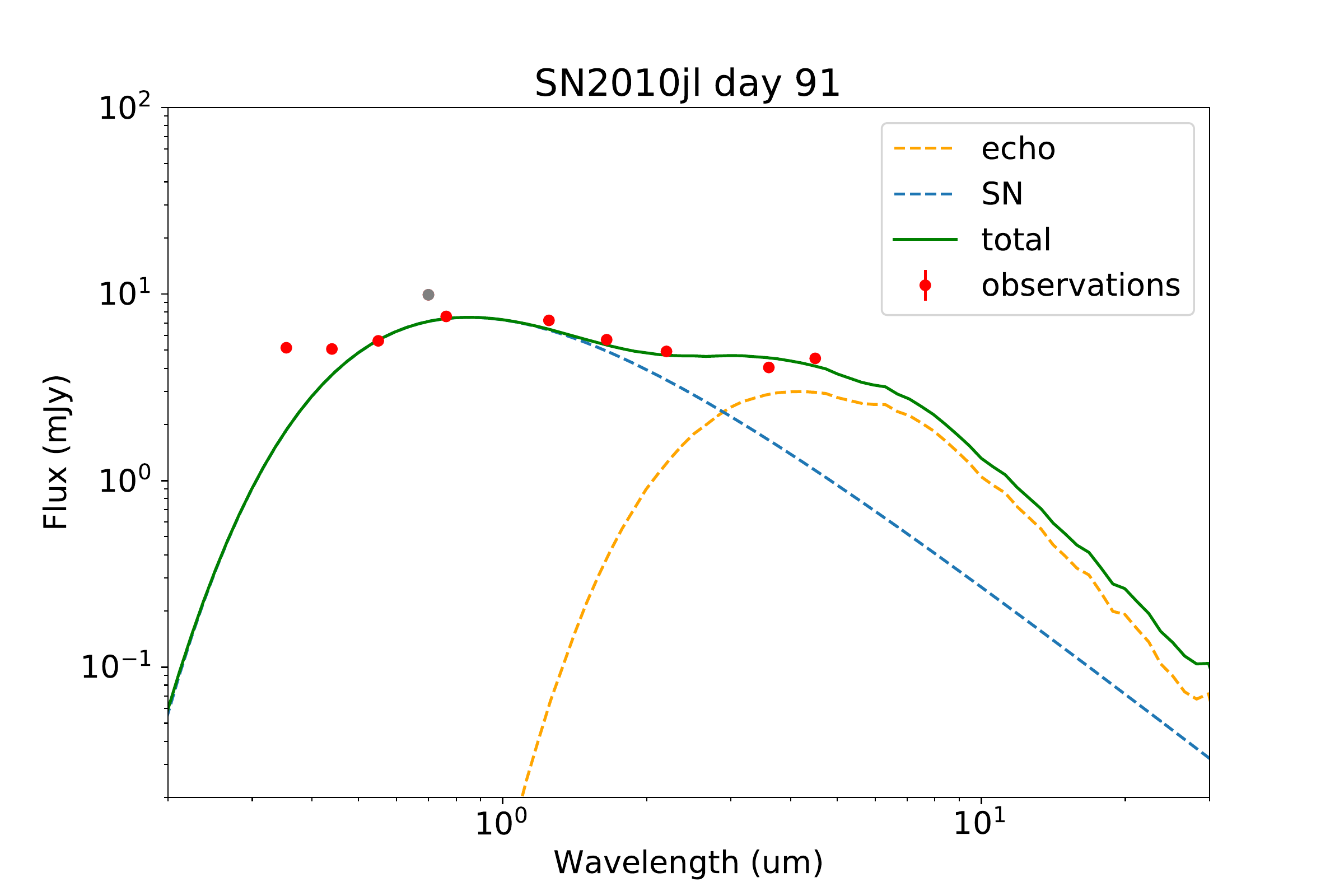}
	\includegraphics[width=0.49\textwidth,clip=true,trim = 23 6 60 35]{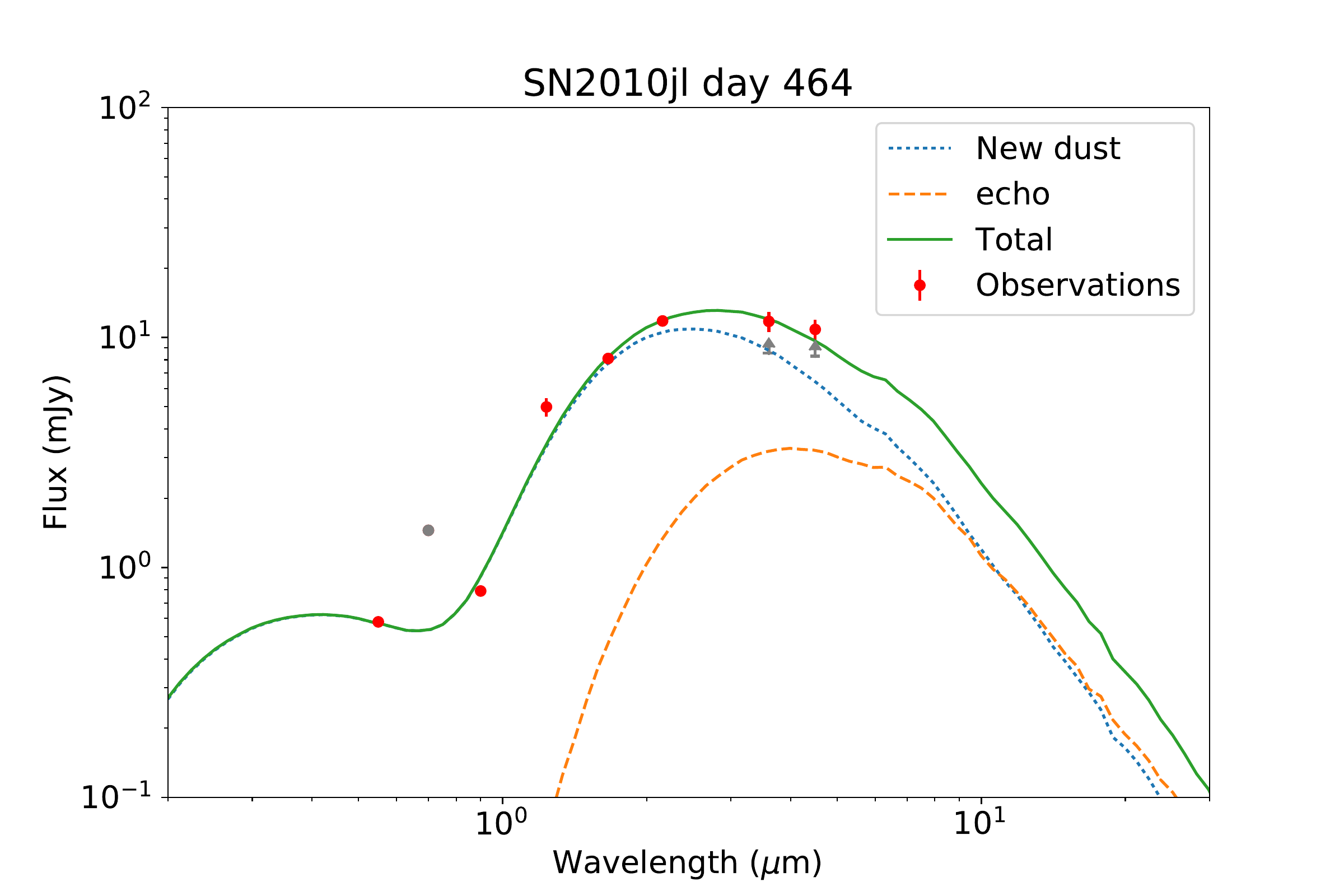}
	\includegraphics[width=0.49\textwidth,clip=true,trim = 23 6 60 25]{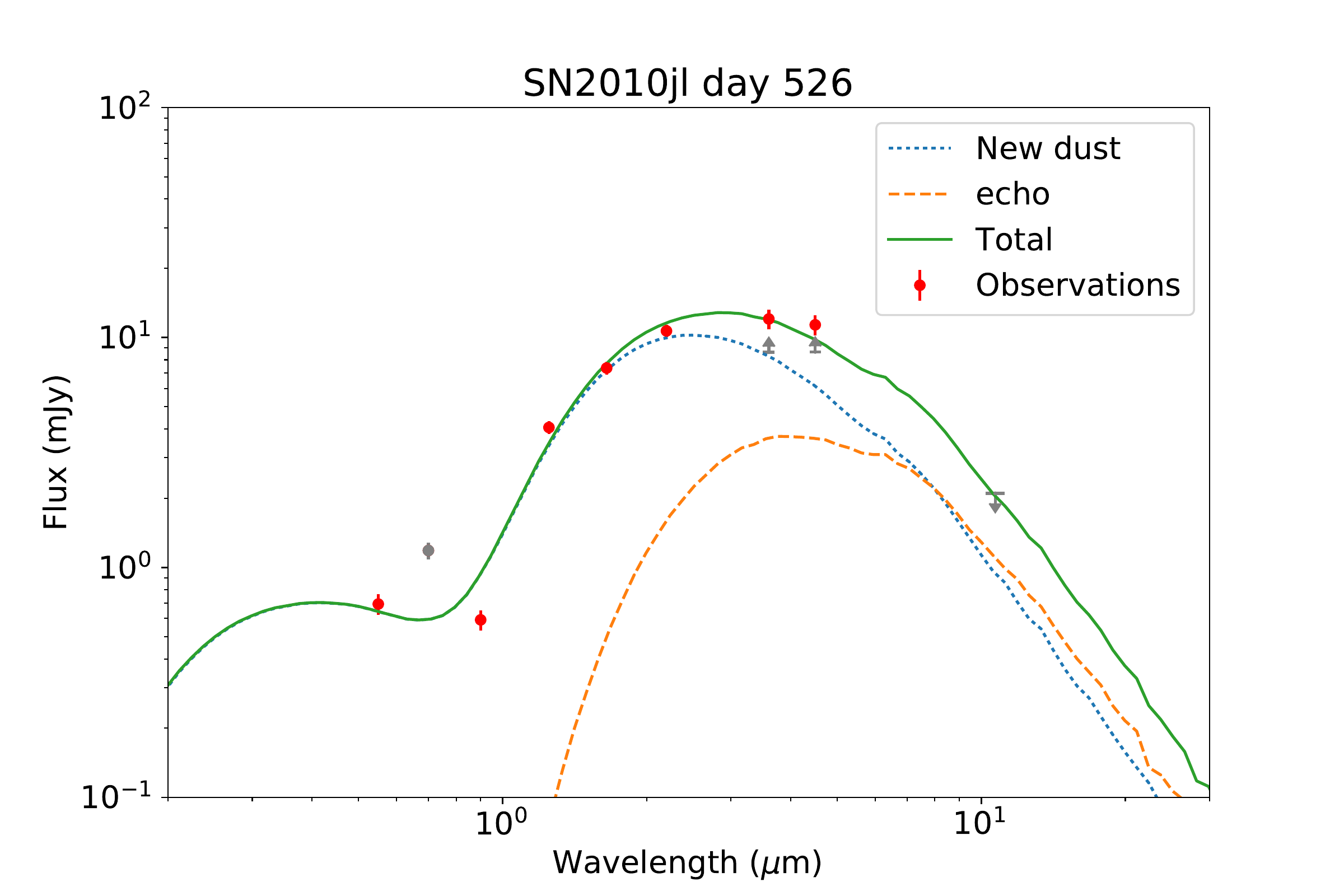}
	\includegraphics[width=0.50\textwidth,clip=true,trim = 10 6 60 25]{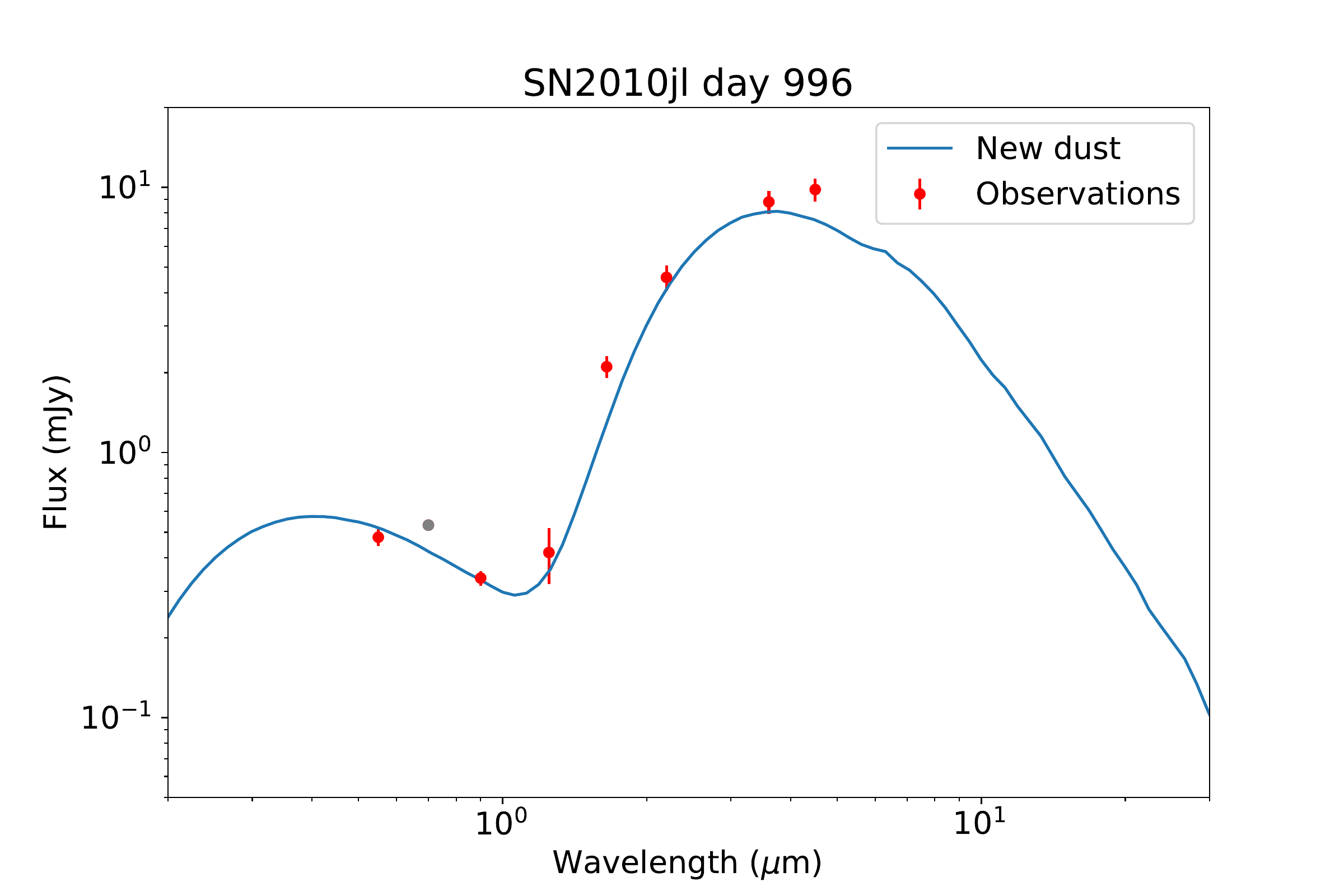} 
	\includegraphics[width=0.50\textwidth,clip=true,trim = 10 6 60 25]{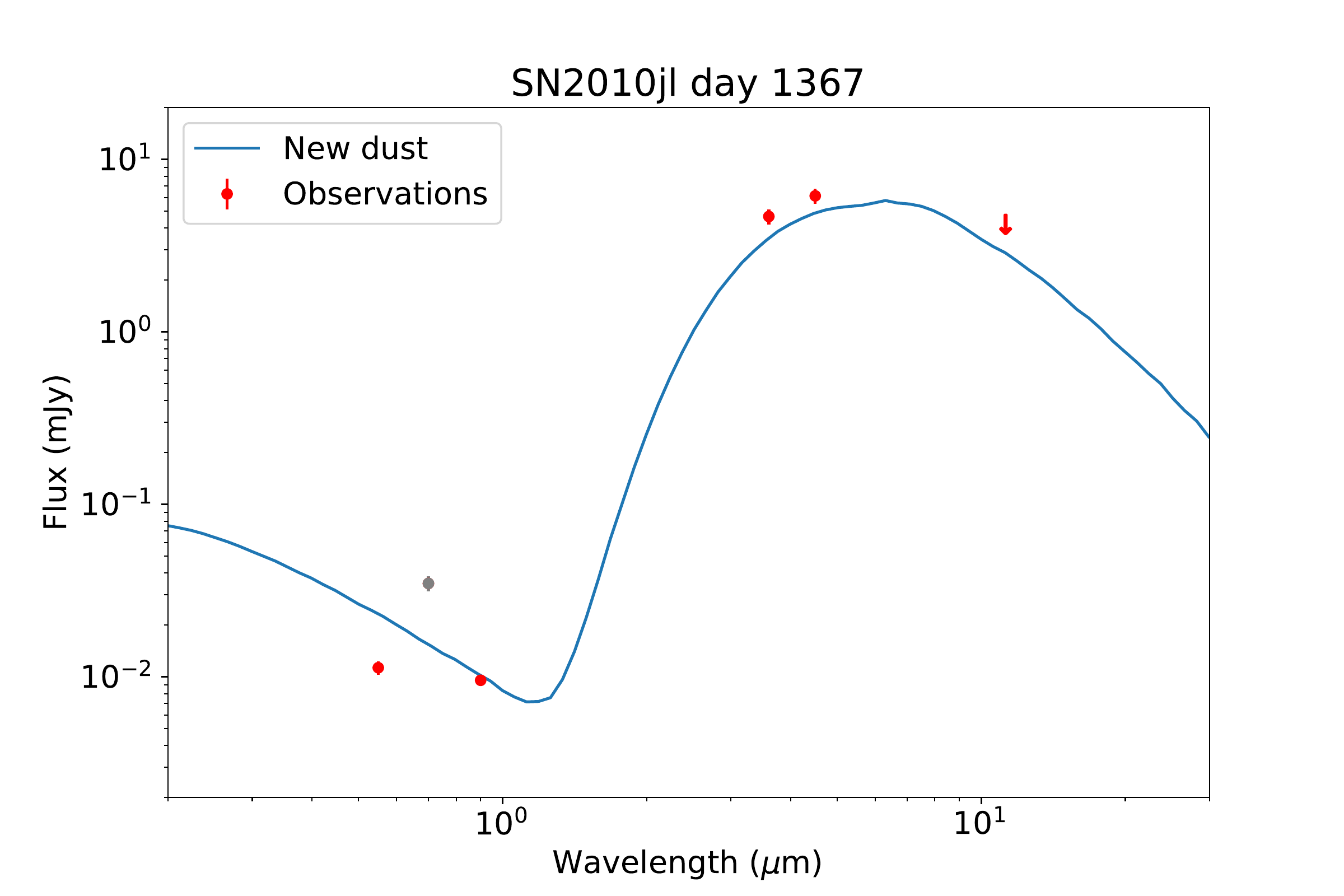} 
	\caption{The results of the {\sc{mocassin}} models, with data points in red. The optical and JHKs points are a combination of our data and that of \citet{2014ApJ...797..118F}, where we used data from the days closest to our mid-IR observations. We do not attempt a fit to the R-band data as they contain the strong H$\alpha$ emission line. The saturated Spitzer fluxes at 464\,d and 526\,d are shown as a lower limit in grey. The VISIR 10.7$\mu$m upper limit at 526\,d is also shown in grey. {\it Top left:} the fit to day 91 showing that the IR can be accounted for by just the flash echo (orange). {\it Top right:} The fit to day 464 shows that the echo is still relevant (orange), but the IR emission is dominated by the dust in the ejecta/CDS (blue). The sum of these components is shown in green. {\it Middle left:} day 526 is very similar to day 464. {\it Middle right:} by day 996 the supernova flash is no longer illuminating the CSM that was echoing at day 91, and the IR emission is coming entirely from new dust (blue).
	{\it Bottom:} Similar to day 996, but with a constraining upper limit at 11.1\,\micron.  }
	\label{Fig:mocassin}
\end{figure*}

We ran a series of models for the later epochs (at 464, 525, 996 and 1367\,d) in which we put dust inside the forward shock radius ($R_{\rm out}$) with a $r{^{-2}}$ density distribution and diffuse photon source representing heating from the gas which has itself been heated by high energy radiation from the interaction of the ejecta with the CSM. All models maintained fixed radii, calculated as per Section \ref{sec:geometry} and listed in Table \ref{tb_mocassin}). A 100\% pure amorphous carbon dust composition was used with grains of $a\geq$0.1~\micron\ radius, based on the {\sc Damocles} results. At each epoch, the dust mass, filling factor, source temperature, grain radius and source luminosity were varied to obtain best fits to the observations (excluding the R-band). A summary of the best-fitting parameters is given in Table \ref{tb_mocassin} and the fits are presented in Figure \ref{Fig:mocassin}. 

At day 464, a clumped distribution allowed for better fits than a smooth distribution of dust which did not allow enough optical emission to escape. We adopted a filling factor of $f=0.1$ for consistency with the line profile models. We found that a source luminosity of $\sim2\times 10^{9}$ L$_{\sun}$ was optimal.  The total mass of newly-formed dust for day 464 was $7 \times 10^{-4}$\,M$_{\sun}$. At 519\,d, the 10.7-$\mu$m upper limit from VLT/VISIR does not allow for very much dust emitting at that wavelength, thus additionally constraining the outer radius of the echoing torus. This upper limit also precludes the presence of a significant fraction of silicate grains.

On day 996, newly-formed dust behind the forward shock is much more important than dust in the echoing torus. Whilst a reasonable fit to the SED can be obtained with grains of radius $a=0.1$~\micron, an improved fit can be obtained with grains of radius $a=0.2~$\micron. The mass of newly formed dust at this epoch had increased from $7 \times 10^{-4}$\,M$_{\sun}$ at day 464 to 3$ \times 10^{-3}$\,M$_{\sun}$ at 996\,d. 

By day 1367, the dust mass required to fit the SED was $1.0 \times 10^{-2}$\,M$_{\sun}$ using grains of radius $a=0.2~$\micron.  A smooth distribution produced a slightly improved fit, but the clumped distribution, which is also a good fit, is consistent with the {\sc{damocles}} results.

\section{Discussion}
\label{sec:discussion}
\begin{figure*}
	\centering
	\includegraphics[width=0.49\textwidth,clip = true, trim = 18 10 42 44]{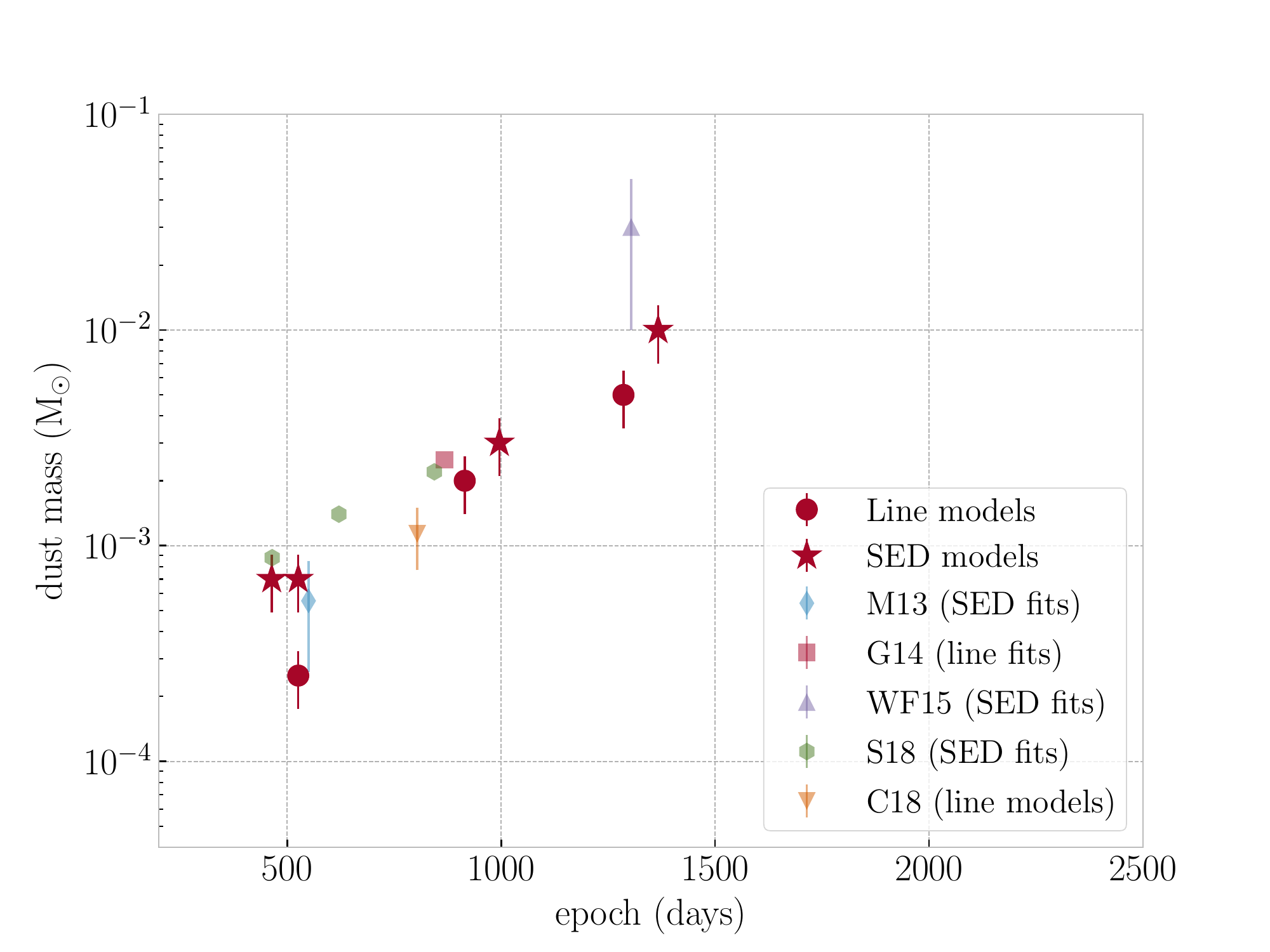} 
    \includegraphics[width=0.49\textwidth,clip = true, trim = 18 10 42 50]{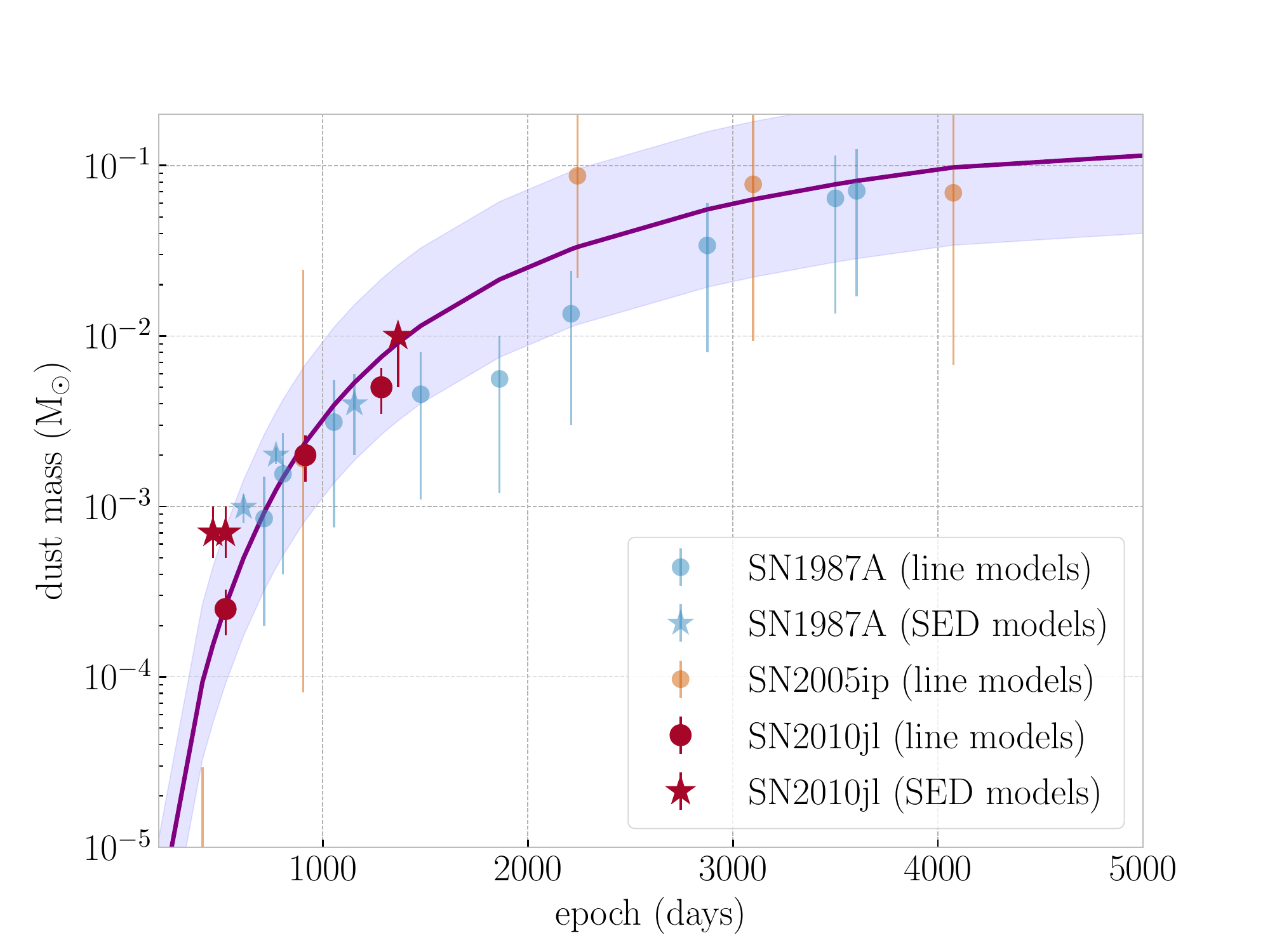}
	\caption{The dust mass (M$_{\sun}$) evolution of SN2010jl. The results of our {\sc{mocassin}} and {\sc{damocles}} modeling are shown above in red stars and circles, respectively. On the left are additionally plotted dust mass estimates for SN 2010jl from  \citet{2013ApJ...776....5M,2014Natur.511..326G,2015ApJ...808L..22W,2018ApJ...859...66S,2018MNRAS.481.3643C}, labelled as M13, G14, WF15, S18, and C18 respectively. On the right, dust mass estimates for SN 2005ip from line profile models \citep{2019MNRAS.485.5192B} and for SN 1987A  from line profile models \citep{2016MNRAS.456.1269B} and SED models \citep{2015MNRAS.446.2089W} are also plotted along with the curve of best fit from \citet{2019MNRAS.485.5192B}.}
	\label{Fig:dustmass}
\end{figure*}

Our radiative transfer models fitting both dust emission and extinction in SN 2010jl have allowed us to determine dust masses at a range of epochs. By fitting the dust emission in the near- and mid-IR, we were able to determine the dust mass emitting in the CSM, CDS and ejecta in SN 2010jl. 
We have been able to construct a scenario  for SN 2010jl whereby the SN ejecta is interacting with a clumpy CSM that is consistent with both photometric and spectroscopic observations over several years. 
We have obtained extremely good fits to the line profiles and SEDs at a range of epochs.	We group our models into four epochs for the sake of comparing our SED and line profiles results for SN 2010jl on days 91, 464 \& 526, 996 \& 915, and 1286 \& 1367. Figure~\ref{Fig:dustmass} summarizes our results and compares them to those from other studies of SN~2010jl and with other supernova dust mass estimates. 

Our dust mass estimates using {\sc{mocassin}} for the first epoch (day 91) reveal that the IR emission can be entirely accounted for by  flash-heated pre-existing dust in an outer torus of circumstellar material. This is in agreement with the findings of \citet{2011AJ....142...45A}, with a slightly lower outer radius for the torus. Assuming that the total gas mass of the CSM is 10 M$_{\sun}$ \citep{2014ApJ...781...42O,2018ApJ...859...66S} and a gas-to-dust ratio of 200, there should be $\sim$0.05 M$_{\sun}$ of pre-existing echoing dust \citep{2007ApJ...663..866D}. Our {\sc{mocassin}} estimate of the echoing dust mass at this epoch is 0.015 M$_{\sun}$, which is entirely pre-existing dust, consistent with the above mass estimate considering the large uncertainties in the CSM mass, the gas-to-dust ratio and the location of the circumstellar material. Some  pre-existing CSM dust may have been vaporized by the initial SN flash, but the early IR emission indicates that a significant mass of CSM dust survived, at least at the radius of the outer torus ($\sim$0.7\,ly). This pre-existing dust associated with pre-supernova mass loss phases would have been more likely to survive the flash if it was distributed in an asymmetric geometry which provided a high dust optical depth along some lines of sight, e.g. a torus or bipolar geometry, as we have inferred here and in \citet{2011AJ....142...45A}. 

The data for the second epoch were obtained at 464 days for {\it Spitzer} and at 526 days for the optical spectra and photometry. As shown in Figure~\ref{Fig:lightcurve}, there is a gap in the data from 260-373 days. After this gap, the SN had become much brighter in the near- and mid-IR, and the red-blue emission line asymmetries had strengthened, indicating the presence of dust in the ejecta and/or CDS (see Figure~\ref{Fig:balmer}). Also during this period, the visible light curve begins a steeper decline. This drop in the optical could be attributed to a number of causes \citep[see][]{2019PASP..131e4204O}, but its coincidental timing with the sharp rise in the IR and increasing line asymmetries may indicate a link with the onset of dust formation in the CDS or ejecta. The first observations after the gap were on day 374 (JHKs) and day 464 ({\it Spitzer}). There is a much larger IR excess at HKs and at 3.6 and 4.5 \micron~than was observed before the gap. Our models reveal that a small contribution from pre-existing, echoing dust in the CSM (see Figure \ref{Fig:mocassin}) still persists at day 464. The {\sc{mocassin}} model on day 464 also requires new  dust with a mass of $8\times10^{-4}$ M$_{\sun}$. The first dust mass estimate using {\sc{damocles}} is 2.5 $\times$ 10$^{-4}$ M$_{\sun}$ for day 526. We discuss the small discrepancy between these dust mass estimates in more detail below. 

The data for the third epoch were obtained at 996 days for {\it Spitzer} and at 915 days for the optical spectra. {\sc Mocassin} models were able to constrain the dust grain radius at this epoch to $\sim$0.2~\micron. Consistent line profile fits were obtained using {\sc Damocles} using this grain radius. The SED models yielded a mass of new dust of $3 \times 10^{-3}$  M$_{\sun}$, whilst at a similar epoch (day 915) the {\sc{damocles}} models required a dust mass of $2\times10^{-3}$  M$_{\sun}$. These masses are in good agreement with each other suggesting that the SED is likely tracing all new dust at this epoch.
 \citet{2014Natur.511..326G} estimated a dust mass of $\sim $ 10$^{-3}$  M$_{\sun}$ at 868 days. This estimate depends on emission out to just the Ks-band and so is not sensitive to cooler dust emitting in the {\it Spitzer} bands and so is broadly consistent with out estimates here. Their inference that larger dust grains are required to account for the line profiles is also in agreement with our conclusions that single-size grains of radius $\sim$0.2~\micron\ are required. Whilst they require a much larger maximum grain size of 4.2~\micron, their use of a power-law distribution of grains steepening towards smaller grains makes direct comparison with our single grain size difficult. 

In our final epoch (1367 days for our SED models and 1286 days for the optical spectra) the ejecta has now reached $\sim 2 \times$10$^{17}$ cm and is significantly mixed with shocked CSM clumps. A slightly improved fit to the photometric observations could be obtained using a smooth distribution instead of simply expanding the model used at previous epochs with a clump volume filling factor of 0.1. However, the {\sc{damocles}} models are inconsistent with a smooth dust distribution at this epoch, requiring clumping to suppress the extended red scattering wings produced by larger grains, which are themselves required to reproduce the observed wavelength dependence of the line asymmetries. The mass of dust inferred from {\sc {damocles}} models reaches 5 $\times$ 10$^{-3}$ M$_{\sun}$ by day 1286. The total warm dust mass of SN 2010jl based on the {\sc{mocassin}} models has also increased, reaching 1 $\times$ 10$^{-2}$ M$_{\sun}$ on day 1367. A plot of the newly-formed dust mass evolution over time, as determined by our SED and line profile fits, is presented in Figure~\ref{Fig:dustmass}.

\subsection{The location of the dust in SN 2010jl}

 Using two independent methods, we have traced the dust in SN 2010jl yielding broadly consistent results.  However, a small discrepancy is seen between the methods, with SED-derived dust masses being slightly higher than those derived from line profiles. These differences are within the uncertainties, but we discuss possible physical causes below. 
	
A limitation of modelling the line profiles of interacting supernovae is the relative extents of the  line emitting region and the unshocked ejecta. Since the projected area of the unshocked ejecta is small compared to the line source, it is possible to hide significantly large masses of dust in the unshocked ejecta that do not significantly affect the line profiles. In the case of SN 2010jl,  our models showed that large masses of dust in the unshocked ejecta would have only a very small effect on the shape of the line profiles.  We therefore emphasise that the masses of dust inferred from our line profile models of SN 2010jl must be located in the post-shocked region. 

Our SED models, adopting the same geometry as our line profile models,  required slightly higher dust masses to reproduce the observed SED. These models were somewhat insensitive to the inner radius of the emitting, dusty shell. Reducing the inner radius to much smaller values of $R_{\rm in}$ than given above did not alter the dust mass required to fit the SED. This new geometry, with smaller inner radii, could be interpreted as a single shell representing both the post-shock region and the unshocked ejecta, though clearly a single power-law density distribution representing both regions is likely a significant simplification of the true geometry. Nonetheless, it suggests the possibility that the emitting dust traced by our SED models could be newly-formed warm dust in the unshocked ejecta or dust in the post-shocked region (or a combination). This may account for the discrepancy with the line profile models which traced only CDS dust.  

We note the theoretical possibility that the presence of cold dust in the unshocked ejecta radiating in the FIR could evade both techniques. 
Whilst we have confirmed that the line profiles would indeed be insensitive to such cold dust, the intense radiation emitted by the reverse shock would heat (or evaporate) any dust in the unshocked ejecta causing it to appear in the SED. Only in cases of extreme clumping might it be possible for cold dust to avoid being heated. The possibility of ejecta geometries that could plausibly hide significant cold dust masses will be the subject of a future paper (Bevan \& Wesson in prep.). We conclude that the total mass of newly-formed dust within SN 2010jl cannot be significantly higher than our dust mass estimates.

\subsection{The origin of the dust in SN 2010jl}

Throughout this work, we have attributed the presence of dust in the post-shocked region of SN 2010jl to newly-formed dust that has condensed in rapidly cooling regions from either shocked ejecta or shocked CSM . An alternative explanation for the presence of dust in the post-shocked region of SN 2010jl at later times is that pre-exsiting CSM dust was overrun by the forward shock but survived its passage to enter the post-shocked region. However, for the adopted torus geometry, our models at 90\,d preclude the presence of dust at radii $\lesssim 0.7$\,ly. Pre-existing dust at smaller radii is heated to higher temperatures than dust further from the illuminating flash. The NIR and MIR SED can therefore constrain the location of any flash-heated, pre-existing dust and our 90\,d model ruled out dust interior to $\sim$0.7\,ly. By $\sim$1300\,d a forward shock travelling at 15,000\,km\,s$^{-1}$ would have reached a distance of 0.18\,ly.	The lack of dust at radii $\lesssim$0.7\,ly at 90\,d is therefore inconsistent with this explanation. We infer that new dust formation interior to the forward shock is required to account for the dust present in SN 2010jl at later times.

\subsection{Comparison with other dust mass estimates}

Several previous works have also concluded that new dust is forming in SN~2010jl, with a range of estimated dust masses in various locations. We compare our dust mass estimates for SN 2010jl with those from the literature in the left panel of Figure \ref{Fig:dustmass}. 

\citet{2013ApJ...776....5M} estimated $7.5-8.5\times10^{-4}$ M$_{\sun}$ of new dust was present in SN2010jl at 550\,d post-outburst, which started forming $\sim$1 year post-explosion. Their analysis was based on the NIR SED, balancing the mass of dust required to produce the emission in the NIR with the absorption required to reproduce the optical data. As might be expected, this is in agreement with our day 464 {\sc mocassin} dust mass estimate but, like our own SED-derived dust mass estimate, is somewhat larger than the dust mass inferred from our line profile modelling at day 526. \citet{2013ApJ...776....5M} further concluded that the dust was likely located in clumps with a filling factor of $\sim$0.1, consistent with our model assumptions. Similarly, \citet{2012AJ....143...17S} inferred that dust may have formed before 500\,d based on the presence of an early IR-excess and the wavelength-dependent asymmetries exhibited by the optical and NIR emission lines. 

 \citet{2014Natur.511..326G} conducted a detailed study of the extent of the red-blue asymmetries in the optical and NIR emission lines concluding that, for a power-law grain radius distribution,  the maximum grain radius must be large (4.2 $\mu$m) to account for the observed wavelength dependence. Our line profile models are restricted to a relatively narrow wavelength range (4861 - 6563\AA) over which we see little variation in the dust optical depth. Given the uncertainties involved and the narrow wavelength range investigated in this work, we do not conclude that there is any conflict with the conclusions of \citet{2014Natur.511..326G} or \citet{2012AJ....143...17S} who highlight strong wavelength dependence in the asymmetries of the optical and NIR line profiles. \citet{2014Natur.511..326G} proposed a rapid (40 to 240 days) formation of carbonaceous dust in the CDS. At later times (500 to 900 days), they concluded that the dominant dust emission component has transitioned to the ejecta. By 868\,d, their inferred dust mass was $2.5 \times 10^{-3}$ M$_{\sun}$, in strong agreement with our dust mass estimate of $2 \times 10^{-3}$  M$_{\sun}$ inferred from our line profile fits, and $3 \times 10^{-3}$  M$_{\sun}$ inferred from our SED fits. 

 \citet{2018MNRAS.481.3643C} conducted a similar modelling of the optical and NIR line profiles of SN 2010jl, reaching similar conclusions as to the origins of the BWC in the rapidly expanding ejecta and the IWC in dense CSM clumps. Though he restricted his analysis to studies of the necessary optical depths and albedos required to reproduce the line profiles, he suggested that, for a grain radius distribution 0.001~\micron\
 $<a<0.1~$\micron\ with $n(a)\propto a^{-3.5}$,  $1.5 \times 10^{-3}$  M$_{\sun}$ of graphite dust in the unshocked ejecta and CDS would be required to fit the line profiles at 804\,d post-outburst, again in broad agreement with out conclusions at similar epochs.  

 \citet{2018ApJ...859...66S} suggested that the IR emission observed prior to 380\,d must be the product of a light echo since dust would not be able to form during the first year of SN~2010jl's evolution. We did not investigate new dust formation during this period. However, we were not able to fit the SED at 91\,d with a pure ejecta dust model and find, in agreement with \citet{2018ApJ...859...66S}, that a pure echo model is able to entirely account for the observed photometry. Whilst the contribution from the echo is still important at 464\,d, the dominant dust emission component has transitioned to the ejecta/CDS by this time.  \citet{2018ApJ...859...66S} estimated interior dust mass of $8.8 \times 10^{-4}$  M$_{\sun}$, $1.4 \times 10^{-3}$  M$_{\sun}$ and $2.2 \times 10^{-3}$  M$_{\sun}$ at 465\,d, 621\,d and 844\,d respectively for an amorphous carbon composition. These estimates closely follow our inferred dust mass estimates.
 
 We compare our dust mass estimates for SN 2010jl to dust mass estimates obtained using detailed radiative transfer models for other core-collapse supernovae in Figure \ref{Fig:dustmass}. We compare SN 2010jl specifically to SN2005ip, as a much less luminous but still dusty Type IIn supernovae for which the dust evolution has been previously investigated \citep{2019MNRAS.485.5192B}. We also compare our results to the dust evolution of SN~1987A as the best-known and most analysed dust-forming core-collapse supernova \citep[e.g.][]{2015MNRAS.446.2089W,2016MNRAS.456.1269B}. We plot the curve of best fit derived for SN2005ip by \citet{2019MNRAS.485.5192B}. There is strong agreement between the dust mass estimates derived for all three of these objects, and similar trends in their dust mass evolution are observed. This is perhaps surprising given differences in their intrinsic types, progenitors and progenitor masses. We conclude that SN 2010jl is likely to continue forming dust in its ejecta/CDS and that future spectral and photometric observations of SN 2010jl will help to determine its final dust mass.

\section{Conclusions}
\label{sec:concs}
Through the use of two complementary radiative transfer codes, we have been able to determine the masses and evolution of different dust populations in SN 2010jl over the first 1400 days. We find that an IR echo is able to account for the observed IR excess at 91\,d but that low line-of-sight optical depths are required. We find that the IR echo makes only a small contribution to the observed IR flux by 464\,d, and makes no contribution by 996\,d. The presence of dust behind the forward shock is required to account for both the blueshifted optical emission line profiles and the observed IR-excess. We have modelled both of these observational signatures of dust in SN 2010jl and have found a dust mass that increases from $2.5-7\times 10^{-4}$\,M$_{\sun}$ at $\sim$500\,d to $0.5-1\times10^{-2}$\,M$_{\sun}$ by $\sim$1400\,d. We have compared our results to the conclusions of other authors and find them to be in generally good agreement. Future observations will be necessary to follow the ongoing formation of dust in SN2010jl.

\acknowledgments

We are grateful for the support provided by the Ball Family Distinguished Professor endowment. This work was also supported by
Spitzer Space Telescope RSA 1439444 and RSA 1415602, issued
by JPL/Caltech, and by European Research Council
(ERC) Advanced Grant SNDUST 694520.
We thank the anonymous referee for their detailed review and constructive suggestions and comments.

\facilities{Gemini (GMOS), VLT (X-SHOOTER), NTT (SOFI), 
{\em Spitzer} (IRAC)}

\software{{\sc iraf}  \citep{Tody1986, Tody1993}; 
	{\sc esorex} \citep{2013A&A...559A..96F}; 
	{\sc mocassin} \citep{2003MNRAS.340.1136E};
	{\sc damocles} \citep{2016MNRAS.456.1269B,2018ascl.soft07023B}
}
\vspace{0.1cm}

\bibliography{everything3}
\bibliographystyle{aasjournal}
\end{document}